\documentclass[twocolumn,showpacs,preprintnumbers,amsmath,amssymb]{revtex4}
\usepackage{epsfig}

\usepackage{graphics}
\usepackage{array}
\usepackage{hhline}
\usepackage{subfigure}
\usepackage{euscript}
\newcommand{\be}{\begin{equation}}
\newcommand{\ee}{\end{equation}}
\newcommand{\bea}{\begin{eqnarray}}
\newcommand{\eea}{\end{eqnarray}}
\newcommand{\nn}{\nonumber\\}
\begin{document}

\title{Dilepton production by dynamical quasiparticles
in the strongly interacting \\
quark gluon plasma}
%\title{Dilepton production by virtual quarks and gluons
%in the strongly interacting \\
%quark gluon plasma}

\author{O.~Linnyk}
\email{linnyk@fias.uni-frankfurt.de}

\affiliation{%
 Institut f\"ur Theoretische Physik, %
 Johann Wolfgang Goethe University, %
%% Max-von-Laue-Str. 1, %
 60438 Frankfurt am Main, %
 Germany %
}

\date{\today}

\begin{abstract}
The dilepton production by the constituents of the strongly
interacting quark-gluon-plasma (sQGP) is addressed.
%
%EITHER: Taking into account not only the higher order pQCD reaction mechanisms,
%but also the non-perturbative spectral functions and self-energies
%of the quarks, anti-quarks and gluons
%thus going beyond the leading twist permits the application
%at realistically low temperatures ($O(T_c)$),
%experimentally relevant low dilepton mass ($O(1\mbox{ GeV})$)
%and strong coupling ($\alpha_S \! \sim \! 0.5 \! \div \! 1$).
%
%OR:
In order to make
quantitative predictions at realistically low plasma temperatures
($O(T_c)$), experimentally relevant low dilepton mass ($O(1\mbox{ GeV})$)
 and strong coupling ($\alpha_S \! \sim \! 0.5 \! \div \! 1$), we
 take into account not only the higher order pQCD reaction
 mechanisms, but also the non-perturbative spectral functions
% (off-shellness)
and self-energies of the quarks, anti-quarks and gluons
thus going beyond the leading twist.
For this purpose, our
calculations utilize parametrizations of the non-perturbative
propagators for quarks and gluons provided by the dynamical
quasi-particle model (DQPM) matched to reproduce lattice data. The DQPM
describes QCD properties in terms of single-particle Green's
functions (in the sense of a two-particle irreducible approach) and
leads to the notion of the constituents of the sQGP being effective
quasiparticles, which are massive and have broad spectral functions
(due to large interaction rates).
In the present work, we derive the off-shell cross sections of
dilepton production in the reactions $q+\bar q\to l^+l^-$ (Drell-Yan
mechanism), $q+ \bar q\to g+l^+l^-$ (quark annihilation with the
gluon Bremsstrahlung in the final state), $q(\bar q)+g\to q(\bar q)+
l^+l^-$ (gluon Compton scattering), $g\to q+\bar q+l^+l^-$
and $q(\bar q)\to q(\bar q)+g+l^+l^-$ (virtual gluon decay,
virtual quark decay) in the sQGP by dressing the quark and gluon lines in the
perturbative diagrams with the DQPM propagators for quarks and
gluons.
%
%The gauge independence of the rates was checked by studying the
%cross sections as functions of the gauge parameters.
%
\end{abstract}


\maketitle

%%%%%%%%%%%%%%%%%%%%%%%%%%%%%%%%%%%%%%%%%%%%%%%%%%%%%%%%%%%%%%%%%%%%%%%%%%%%%%%%%%
%%%%%%%%%%%%%%%%%%%%%%%%%%%%%%%%%%%%%%%%%%%%%%%%%%%%%%%%%%%%%%%%%%%%%%%%%%%%%%%%%%
\vspace{10pt}
\section{Introduction} \label{intro}

\begin{figure*}%[b]
\begin{center}
\resizebox{\textwidth}{!}{%
 \includegraphics{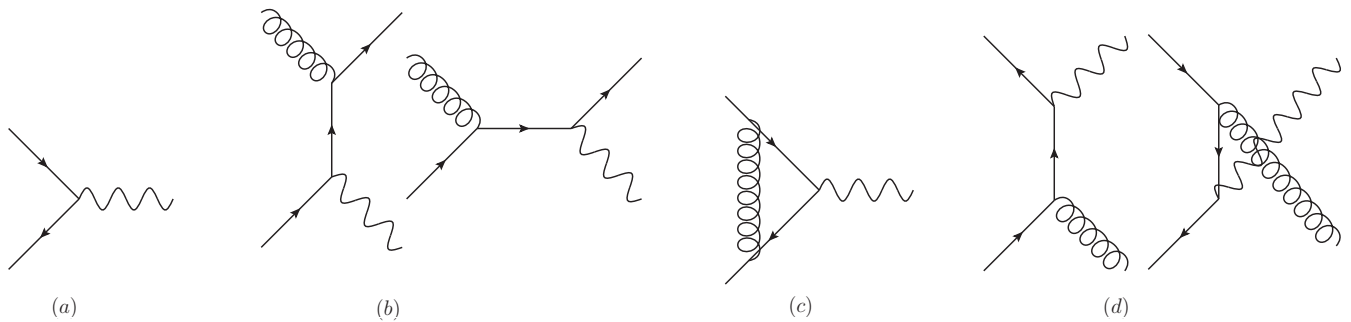}
} \caption{Perturbative QCD diagrams contributing
to the dilepton production up to the order $O(\alpha_S)$:
(a) Drell-Yan mechanism, (b) gluon Compton scattering (GCS),
(c) vertex correction, (d) quark annihilation with gluon Bremsstrahlung.
Virtual photons (wavy lines) split into lepton
pairs, spiral lines denote gluons, arrows denote quarks. In each
diagram, the time runs from left to right.} \label{diagrams}
\end{center}
\end{figure*}

Since many years the transition between the hadronic phase and the
quark-gluon plasma (QGP) as well as the nonperturbative properties
of QGP motivate a large community and justify large-scale
experiments, in which heavy nuclei are collided at relativistic
energies in order to achieve the high energy-densities necessary for
the transition to the deconfined state of matter. Electromagnetic
probes (i.e. dileptons and photons) are powerful tools to explore
the early (hot and dense) stage of the heavy-ion collision, since,
unaffected by the final state interaction, they carry to the
detector information about the conditions and properties of the
environment at the time of their production -- encoded in their mass
and momentum distributions, -- thus providing a glimpse deep into
the bulk of the strongly interacting
matter~\cite{Tserruya:2009zt,Fries:2002kt}. In 1978, E.~Shuryak
proposed to use dilepton as probes of QGP, after the suggestion was
made that the dileptons and photon yields reflect the properties of
the medium created in hadron-hadron collisions (see the pioneering
works~\cite{Shuryak:1977ut,Shuryak:1978ij,Feinberg:1970tg,Feinberg:1976ua,Bjorken:1975dk}).

Real and virtual photons, i.e. dileptons, are emitted over the
entire space-time evolution of the heavy-ion collision, from the
initial nucleon-nucleon collisions through the hot and dense phase
and to the hadron decays after freeze-out. This is both a challenge
and advantage of electromagnetic probes. Fortunately, lepton pairs
possess an additional degree of freedom (the invariant mass $Q^2$),
which allows to separate different ``physics" by observing the
dilepton radiation in different mass ranges. The low mass
($Q<1$~GeV) spectrum of dileptons -- generated in heavy ion
collisions -- is dominated by the vector meson decays, the
production of lepton pairs of high mass ($Q>3$~GeV) is governed by
the perturbative quantum chromodynamics (pQCD), while the dilepton
yield in the intermediate mass range ($1<Q<3$~GeV) is sensitive to
the possible formation of a QGP.

Dilepton measurements have possibly provided a sign of the
deconfined matter at SPS energies.
%The intriguing recent finding of the
The NA60
Collaboration~\cite{ARNALDI:2006JQ,SEIXAS:2007UA,DAMJANOVIC:2007QM}
has recently found that the effective temperature of the dileptons
in the intermediate mass range is lower than the $T_{\mbox{eff}}$ of
dileptons at lower masses, which are of hadronic origin. This can be
explained, if one assumes that the spectrum at the invariant masses
above 1~GeV is dominated by the partonic channels in the
QGP~\cite{Rapp:2009yu,Linnyk:2009nx,Linnyk:2010ar}. In this case,
the softening of the transverse mass spectrum with growing invariant
mass implies that the partonic channels occur dominantly before the
collective radial flow has developed. The assumption that the
dilepton spectrum at masses above 1~GeV could be dominated by QGP
radiation was supported by the studies within the
Hadron-String-Dynamics (HSD) transport approach~\cite{Cass99}, which
have shown~\cite{Bratkovskaya:2008bf} that the measured dilepton
yield at low masses ($Q\le1$~GeV) can be explained by the dilepton
production in the hadron interaction and decay, while there is a
discrepancy between the HSD results and the data in the mass range
above 1~GeV. This access seen at $Q>1$~GeV is not accounted for by
hadronic sources in HSD  -- in-medium or free -- and might be seen
as a signal of partonic matter, manifest already at 158~AGeV
incident energy.

%In 2008,
Recently, the PHENIX Collaboration has presented first dilepton data
from $pp$ and $Au+Au$ collisions at Relativistic-Heavy-Ion-Collider
(RHIC) energies of $\sqrt{s} =
200$~GeV~\cite{TOIA:2005VR,TOIA:2006ZH,AFANASIEV:2007XW,ADARE:2009QK}.
The data show an even larger enhancement over hadronic sources in
$Au+Au$ reactions
%(relative to $pp$ collisions)
in the invariant mass regime from 0.15 to 0.6~GeV, which could not
be explained in the scope of the HSD approach neither by meson
decays -- in-medium or free -- nor by hadronic
Bremsstrahlung~\cite{Bratkovskaya:2008bf}. It is of interest,
whether the excess at RHIC is due to the dominance of sources in the
QGP~\cite{Bratkovskaya:2008bf}.

Early predictions of the dilepton emission from QGP relied on
perturbative formulae for the cross sections of the virtual photon
production in $q+q$ and $q+g$
collisions~\cite{Shuryak:1978ij,Halzen:1978et,Lin:1999uz}. Indeed,
first concepts of the QGP were guided by the idea of a system of
partons which interact weakly, with pQCD cross sections. However,
most theoretical estimates of the temperatures, which are reasonably
expected to be currently achieved in heavy ion collisions are not
extremely large compared to the QCD scale
$\Lambda_{QCD}$~\cite{McLerran:1984ay}. Therefore, the QCD coupling
$\alpha_S$ is not small. In agreement with this early expectation,
experimental observations at RHIC indicated that the new medium
created in ultra-relativistic Au+Au collisions was interacting
strongly - stronger than hadronic matter. Moreover, in line with
theoretical studies in
Refs.~\cite{Shuryak:2003xe,Thoma:2005uv,Peshier:2005pp} the medium
showed phenomena of an almost perfect liquid of
partons~\cite{QM04PHENIX,QM04STAR,QM04BRAHMS,QM04PHOBOS} as
extracted from the strong radial expansion and elliptic flow of
hadrons~\cite{QM04PHENIX,QM04STAR,QM04BRAHMS,QM04PHOBOS}. Studies
performed in the framework of the lattice regularized
QCD~\cite{Karsch:2001vs} have also shown that the high temperature
plasma phase is a medium of interacting partons which are strongly
screened and influenced by non-perturbative effects even at
temperatures as high as $10 \, T_c$.

Consequently, the concept of perturbatively interacting quarks and
gluons as constituents of the QGP had to be given up. Due to large
running coupling, the next-to-leading order (NLO) gluon-quark
interactions are expected to contribute considerably in addition to
the leading order mechanism of quark-quark annihilation ($q\bar q\to
l^+l^-$) to the QGP radiation spectrum. The importance of the higher
order corrections is long understood~\cite{Halzen:1978rx}. On the
other hand, non-perturbative nature of the sQGP constituents
manifests itself not only in their strong coupling, but also in the
modified spectral densities and self energies, which should be taken
into account in consistent calculations of dilepton production from
the QGP.

A cure can be found in reordering perturbation theory: by expanding
correlation functions in terms of effective propagators and vertices
instead of bare ones~\cite{Kapusta:1991qp}.
%This
A powerful resummation technique was developed by Braaten,
Pisarsky~\cite{Braaten:1990wp} and Wong~\cite{Wong:1992nk}. The
production of dileptons was calculated at leading order in the
effective perturbation expansion in~\cite{Braaten:1990wp}, using as
the effective propagators the bare ones plus one loop corrections
evaluated in the high-temperature
limit~\cite{Pisarski:1988vd,Silin,Klimov,Weldon:1982aq}. In this
approach the singularity of the production cross section -- that
dominates the dilepton rate -- is regularized by the thermal masses
of quarks $m_{th}$ and gluons $m_g$, which are in turn determined by
the one-loop leading order result in the thermal perturbation theory
(HTL). The approach has been extended to the dilepton radiation from
non-equillibrium plasmas in~\cite{Baier:1997td,Strickland:1994rf}.

However, since virtual photon rates need to be evaluated at
temperatures that are not very large compared to $T_c$, it is
important to take values for $m_{th}$, $m_g$ not from the HTL
approximation but from, for instance, a fit of the lattice QCD
entropy by a gas of massive quarks equation of
state~\cite{Blaizot:2005mj,Gelis:2002yw}, as has been done
in~\cite{Cassing:2007nb,Peshier:1994zf}. Alternatively, one might
treat thermal masses in the calculation of the dilepton rates as
phenomenological parameters~\cite{Thoma:1999nm}.

%Lattice gauge theory is currently the approach of choice to access
%the nonperturbative properties of the high temperature phase of
%QCD~\cite{Karsch:2001vs}. Such characteristics of quark-gluon matter
%as the deconfinement phase-transition temperature ($T_c$), the
%equation of state, the running coupling at low renormalization
%scales are now fairly well established due to lQCD calculations.
%
Not long ago, a first attempt appeared to calculate directly on the
lattice the production of dileptons in QGP~\cite{Karsch:2001uw}. The
 suppression at small $Q^2$ observed on the
lattice has attracted a lot of interest, because it is not what one
would expect from (resummed) perturbation theory: The finite thermal
masses would indeed produce a drop of the Born (Drell-Yan) term
$q+\bar q\to\gamma^*$ because of the threshold effect -- as
predicted~\cite{Gorenstein:1989ks} in relation to the cut-off in the
momentum distribution of quarks and confirmed in effective
perturbation theory in the
works~\cite{Cassing:2007nb,Peshier:1994zf,Thoma:1999nm}, -- but
there are higher order processes ($q\bar q\to\gamma^*g$,
$qg\to\gamma^*q$) that have no threshold and would fill the spectrum
at small $Q^2$.
The effect of multiple scattering of the quark in the plasma
(Landau-Pomeranchuk-Migdal
effect~\cite{Landau:1953um,Landau:1953gr,Migdal:1956tc}) on the rate
of $q\bar q \to \gamma^*$ was considered in the
context of the semi-classical approximation % ?
in~\cite{LPM}, where it was shown that the multiple scatterings lead
to the disappearance of the threshold behavior in the Born rate.

In the present work, we pursue a model, in which the (multiple)
interaction of quarks and gluon in sQGP is encoded in their
effective broad spectral functions. The non-zero width of quarks is
related to their strong interaction, which is manifest in the
elastic scattering as well as in the virtual gluon emission. By
dressing the partonic lines in the dilepton production processes
($q+\bar q \to l^+ l^-$, $q+ \bar q \to g+l^+l^-$, $q+g \to
q+l^+l^-$) with effective spectral functions we study the effect of
the partonic interactions in the plasma on their dilepton radiation,
especially in the interesting region of low $Q^2$.

The effective propagators were obtained from lattice data in the
scope of the Dynamical QuasiParticle Model (DQPM)~\cite{Cassing06}.
DQPM describes QCD properties in terms of single-particle Green's
functions (in the sense of a two-particle irreducible (2PI)
approach) and leads a quasi-particle equation of state, which
reproduced the QCD equation of state extracted from Lattice QCD
calculations. According to the DQPM, the constituents of the sQGP
are effective strongly interacting massive partonic quasi-particles
with broad spectral functions, i.e. non-vanishing width.

Strong interaction of partons -- reflected in their broad widths --
leads to higher-twist corrections to the standard pQCD cross
sections~\cite{Linnyk:2006mv}. The influence of the higher twists is
hard to estimate and has been ignored in most calculations so far.
Limiting the calculations to the leading twist is a widely used
approximation at high energies for the following reason: The higher
twists by definition are vanishing in the limit of infinite
invariant mass of the lepton pair~\cite{Vainshtein}. Indeed, one
observes that by calculating next-to-leading order contributions and
refitting the parton distributions
accordingly~\cite{Halzen:1978rx,GRV}, one can already significantly
reduce the discrepancy with the data on the double-differential
Drell-Yan cross section in elementary $p+p$
collisions~\cite{STIRLING:1993GC,HAMBERG:1990NP,ELLIS:1991QJ}
compared to the LO predictions.

However, the power-$Q^2$ suppressed (i.e. higher twist)
contributions can be large at realistic energies and/or in
collective systems with a relatively short interaction length like
the sQGP created in heavy-ion collisions. It has been
shown~\cite{Linnyk:2006mv} that the contribution of higher twists is
essential for a proper description of the data on the {\em
triple-differential} Drell-Yan cross section. For instance, quark
and gluon off-shellness -- arising due to the non-perturbative
interaction between the partons -- have a large effect on the
transverse momentum distribution of produced lepton
pairs~\cite{Collins:2005uv,Linnyk:2004mt,Linnyk:2006mv}.

In the current paper, we study the dilepton production from
dynamical quasi-particle interactions.
%Indeed, DQPM model captures the
%major details of the nonperturbative properties of the partons in
%the sQGP as calculated on the lattice.
For this purpose we derive the off-shell cross sections of $q+\bar q
\to l^+ l^-$, $q+ \bar q \to g+l^+l^-$, $q+g \to q+l^+l^-$ ($\bar
q+g \to \bar q+ l^+l^-$), $q\to q+g+l^+l^-$ ($\bar q\to \bar
q+g+l^+l^-$) and $g\to q + \bar q +l^+l^-$ by calculating them for
the arbitrary virtualities of external quarks and gluons, while
dressing the internal lines with effective self energies, and take
into account the non-zero widths of partons by convoluting the
obtained cross sections with the effective spectral functions from
the DQPM. Using the DQPM parametrizations for the quark(gluon)
self-energies, spectral functions and interaction strength, we
calculate the dilepton production by partonic interactions in the
sQGP while accounting for their non-perturbative dynamics, including
the higher twist effects.

\begin{figure}
\begin{center}
\resizebox{0.4\textwidth}{!}{%
 \includegraphics{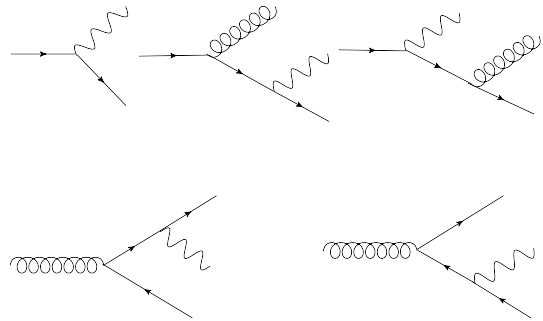}
} \caption{Diagrams contributing
to the dilepton production by virtual quasi-particles
in addition to the ones presented in Fig.~\protect{\ref{diagrams}}.
Upper part: the decay of a virtual quark;
Lower part: the decay of a virtual gluon.
Virtual photons (wavy lines) split into lepton
pairs, spiral lines denote gluons, arrows denote quarks.}
\label{diagrams2}
\end{center}
\end{figure}

The perturbative
diagrams for the dilepton production at order up to $O(\alpha_S)$
are illustrated in
Fig.~\ref{diagrams}. Let us briefly summarize the differences of our
effective theory approach from the standard pQCD:
\begin{itemize}
\item
(a) we take into account full off-shell kinematics, in particular
the transverse motion and virtuality of the partons,
\item
(b) quark and gluon lines are dressed with non-perturbative spectral
functions and self-energies: the cross sections are derived for
arbitrary masses of all external parton lines and integrated over
these virtualities weighted with spectral functions
(see e.g. Refs~\cite{Bratkovskaya:2008iq,Linnyk:2006mv}
for an introduction to the method); the internal
lines are dressed with self energies,
\item
(c) vertices are modified compared to pQCD by replacing the
perturbative coupling (that runs with the momentum transfer) with
the full running coupling $\alpha_S$ that depends on the temperature
$T$ according to lattice data parametrization by~\cite{Cassing06},
while the temperature is related to the local energy density
$\epsilon$ by the lQCD equation of state. Note that close
to $T_c$ the full coupling increases with the decreasing temperature
 much faster than the pQCD prediction.
\item
(d) Due to the broad width for quarks and gluons in the
sQGP~\cite{Cassing:2007nb} -- which is the consequence of their high
interaction rate,-- we obtain non-vanishing contributions also from
the processes of the decays of virtual quarks ($q\to q+g+l^+l^-$)
and gluons ($g\to q + \bar q +l^+l^-$), which are forbidden
kinematically in pQCD (see Fig.~\ref{diagrams2}).
\end{itemize}
Thus, we
utilize the DQPM parametrization for the effective quark and gluon
propagator while simultaneously taking into account the
NLO reaction mechanisms. This way, we are going
in our calculations beyond the leading order in $\alpha_S$
{\em and} depart from the leading twist approximation.

The paper is organized as follows.
The analysis of the off-shell kinematics and the calculations of the
off-shell cross sections in our model are given in
section~\ref{offshellsection} for each of the partonic processes
separately. On the other hand, in the limit of high hard scale
$Q^2$, the off-shell cross sections should approach the perturbative
ones~\cite{jaffe,Vainshtein}. Therefore, let us first
recapitulate the corresponding pQCD results in section~\ref{pQCD}.
%...
The off-shell cross sections will be compared to the perturbative
ones throughout the section~\ref{offshellsection}.

In Section~\ref{relative} we give a simple example of an application
of the off-shell cross sections derived above to the calculation of
the relative contributions of different processes to the sQGP
radiation. We point out, however, that the considerations in the
current paper will probably be not detailed enough to be applied to
the description of the hot and dense early phase of relativistic
heavy-ion collisions, where sQGP is formed. A quantitative
comparison to the experimental data and reliable conclusions on the
relative contribution of various sources to the experimentally
observed thermal dilepton
spectrum~\cite{ARNALDI:2006JQ,SEIXAS:2007UA,DAMJANOVIC:2007QM,TOIA:2005VR,TOIA:2006ZH,AFANASIEV:2007XW,ADARE:2009QK}
requires taking into account the non-equilibrium dynamics of the
heavy-ion collision in its full complexity by use of microscopic
transport models, which is beyond the scope of this study.
The main purpose of the current paper is to built an effective
approach for the derivation of the off-shell cross sections for the
interaction of dynamical quasi-particles as constituents of the sQGP.
The qualitative analysis of the relative importance of different
processes in section~\ref{relative} should be understood as an illustration to
the above results rather than a quantitative
prediction for the dilepton yield from heavy-ion collisions.

Section~\ref{effectWidth} is devoted to analyzing the effect of
finite quark and gluon widths on the dilepton rate. In
Section~\ref{conclusions} we summarize the main results and their
possible applications.
%The important question of the gauge invariance of the obtained rates and
%cross sections is addresses in the appendices.

%%%%%%%%%%%%%%%%%%%%%%%%%%%%%%%%%%%%%%%%%%%%%%%%%%%%%%%%%%%%%%%%%%%%%%%%%%%
\vspace{10pt}
\section{Dileptons from perturbative partons}
\label{pQCD}

In the present Section, we consider the following partonic
mechanisms for dilepton production in pQCD:
\begin{enumerate}
\item the Drell-Yan mechanism of quark annihilation
($q+\bar q\to \gamma^*$),
\item quark + anti-quark annihilation with gluon
Bremsstrahlung in the final state ($q+\bar q\to g+\gamma^*$),
\item Gluon Compton scattering ($q+g\to \gamma^*+q$ and $\bar q+g\to
\gamma^*+\bar q$).
\end{enumerate}

%---------------------------------------------------------------------
\subsection{Drell-Yan mechanism}

%\begin{figure}
%\begin{center}
%\resizebox{0.46\textwidth}{!}{%
% \includegraphics{DYon.eps}
%} \caption{Integrated cross section of $q+\bar q\to\mu^+ +\mu^-$
%at leading order and leading twist of pQCD. }
%\label{DYon}
%\end{center}
%\end{figure}

The leading order pQCD mechanism for the dilepton production in the
partonic phase is the same as for the well known Drell-Yan (DY)
process~\cite{Drell.Yan}: quark and antiquark annihilate into a
lepton pair ($q \bar q \to l^+l^-$), as presented by the diagram (a)
in Fig~\ref{diagrams}. The leading order leading twist (LT) pQCD
result for the cross section of DY dilepton production is
\bea \label{OnshellSigmaLO} \hspace{-0.45cm} \left( \frac{d^3 \hat
\sigma (q\bar q\to l^+l^-) }{ d Q^2 d x_F dq_T^2}
\right)^{\mbox{\scriptsize DY}} _{\mbox{\scriptsize on-shell}} \!
\hspace{-0.3cm} & \!  = \! & \! \frac{4 \pi \alpha^2 e_q^2 }{9 Q^4 }
\frac{x_1 x_2}{x_1 + x_2} \left( 1-x_1 x_2 \right) \nn & &
\hspace{-2.4cm} \times  \delta (q_T^2) \delta \left( Q^2 - x_1 x_2
S_{NN} \right) \delta \left( x_F - \frac{x_2-x_1}{1-x_1 x_2}
\right),
 \eea
where $\alpha$ is the electromagnetic fine structure constant, $e_q$
the fractional quark charge, the subscript ``on-shell" stands for
`leading twist', the lepton pair has invariant mass $Q^2$ and
transverse momentum $q_T$.

%$q_T$ that relates to the scattering angle as
%%
%\be q_T^2=\frac{(s-Q^2)^2}{4s} \sin ^2 \Theta , \ee
%%
% Note that the maximum transverse momentum of the Drell-Yan
%pair $(q_T^2)_{max}$ is fixed by kinematics as
%%
%\be (q_T^2)_{max} = \frac{(s+Q^2- M_R ^2)^2}{4 s} - x_F^2
%\frac{(s-Q^2)^2}{4 s} - Q^2,\ee
%%
%where $M_R^2$ is the minimal invariant mass of the undetected
%remnant.

In collinear pQCD, the off-shellness, mass and transverse momentum
of the annihilating quark and antiquark are neglected, and,
therefore, the incoming parton momenta are related to the momenta of
colliding nuclei as $p_{q (\bar q)}=x_i P_A /A$. In this case, the
parton momentum fractions $x_1$ and $x_2$ are related to the
virtuality and $x_F$ of the produced photon as (cf. delta-functions
in (\ref{OnshellSigmaLO})):
\bea \label{x1x2} Q^2 & = & s =  x_1 x_2 S_{NN}; \\ \label{xFLO} x_F
& = & (x_2-x_1)/(1-x_1 x_2). \eea
Note that the denominator of the $x_F$ definition in (\ref{xFLO}) is
omitted in some works, where an approximate definition $x_F\approx
2p_z/\sqrt{S_{NN}}$ is used instead of $x_F=q_z/(q_z)_{max}$, $s$
denotes the invariant energy for the partonic process, $S_{NN}$ --
for the hadronic one. The kinematical limits for this process are
\be S_{NN}  \ge  Q^2, \mbox{ } |x_F|  \le 1, s=Q^2. \ee

%For an illustration, we plot in Fig.~\ref{DYon}, the on-shell cross
%section introduced above for the dilepton production in $q+\bar q\to
%\mu^+\mu^-$ for $Q=0-4$~GeV. %Note that $\sqrt{s}=Q$.

%---------------------------------------------------------------------
\subsection{Gluon Bremsstrahlung}
\label{cutoffintro}

%\begin{figure}
%\vspace{10pt}
%\begin{center}
%\resizebox{0.46\textwidth}{!}{%
% \includegraphics{gBRon.eps} } \caption{Gluon Bremsstrahlung cross
%sections in the on-shell approximation at different $\sqrt{s}$.}
%\label{sqrt_dependence}
%\end{center}
%\end{figure}

The cross section of the {\em gluon Bremsstrahlung} process $\bar q
q \to g + \mu^+\mu^-$ is~\cite{Altarelli:1977kt,Vainshtein:1976kd}
\bea \label{Altarelli.gBrems} \hspace{-0.4cm} \left( \frac{d^2
\hat\sigma (q\bar q\to g l^+l^-) }{ d Q^2 d \cos \Theta}
\right)^{\mbox{\scriptsize gBr}} _{\mbox{\scriptsize on-shell}}
\hspace{-0.3cm} & = & \frac{8 \alpha ^2 e_q^2 \alpha_S}{27 Q^2}
\frac{s-Q^2}{s^2 \sin ^2 \Theta }  \nn & & \hspace{-1cm} \times
\left( 1 + \cos ^2 \Theta +4 \frac{Q^2 s}{(\hat s-Q^2)^2} \right) \!
\! , \eea
where $s$ is the total energy squared of the colliding partons, and
$\Theta$ is the scattering angle of the outgoing lepton pair with
respect to the incoming quark momentum in the quark center-of-mass
system (CMS). Note that the cross section (\ref{Altarelli.gBrems}) can be
written in terms of the Mandelstam variables $s$, $t$ and $u$
as~\cite{Halzen:1978et}
\bea \label{gBrems.stu} \hspace{-0.3cm} \left( \frac{d^2 \hat\sigma
(q\bar q\to g l^+l^-) }{ d Q^2 d t} \right)^{\mbox{\scriptsize gBr}}
_{\mbox{\scriptsize on-shell}} \! \hspace{-0.4cm}  & =
\hspace{-0.2cm} & \frac{8 \alpha ^2 e_q^2 \alpha_S}{27 Q^2}
\frac{(t-Q^2)^2+(u-Q^2)^2}{s^2 t u}  \nn & & \hspace{-1.2cm} % \times
\phantom{\Theta \left( s+t-Q^2 \right)} \times \delta \left(
s+t+u-Q^2 \right)
\\ \label{gBrems.stu2}
& = \hspace{-0.2cm} & \frac{8 \alpha ^2 e_q^2 \alpha_S}{27 Q^2 s^2}
\left( \frac{t}{u}+\frac{u}{t}+\frac{2 s Q^2}{t u} \right)  \nn & &
\hspace{-1.2cm} %\times
\phantom{\Theta \left( s+t-Q^2 \right)} \times \delta \left(
s+t+u-Q^2 \right) \! \!
,
\eea %
which coincides with the QED cross section for the virtual Compton
scattering up to the color factor and the crossing
transformation~\cite{Peskin:1995ev}.
% $t\leftrightarrow s$
Here we
denote the momenta of the incoming quark and antiquark as $p_1$ and
$p_2$, the momenta of the outgoing gluon as virtual photon as $k$
and $q$,
$ s= (p_1+p_2)^2$, $t=(p_1-q)^2$, $u= (p_2-q)^2$. The
$\delta$-function $ \delta(s+u+t-Q^2)$ reflects the on-shell
condition for the partons: \be p_1^2+p_2^2+k^2=0.  \ee

The collinear divergence of the gluon Bremsstrahlung cross section
for $t\to0$ and $u\to0$ (i.e. $\cos{\Theta}\to\pm 1$) is obvious; a
cut-off $\Lambda^2$ on $t$ can be used in order to regularize it:
\bea t & \le & - \Lambda ^2, \\ t & \ge & -s +Q^2+\Lambda^2. \eea
Since in the CMS of the colliding partons we have
\be t=Q^2-\sqrt{s}q^0+\sqrt{s}|\vec{q}|\cos{\Theta},
\ee %
the corresponding cut-off with respect to $\cos{\Theta}$ is
\be | \cos{\Theta}| \le \frac{( \sqrt{s}q^0-Q^2-\Lambda^2
)}{\sqrt{s}|\vec{q}|}.
 \ee
Another divergence
%at $Q\to0$ is to be regularized by means of
%the Landau-Pomeranchuk resummation
%~\cite{Aurenche:2002wq}. %%%????
%
%The third type of divergence
in the perturbative expression (\ref{gBrems.stu}) is the infrared
(IR) divergence for the energy of the gluon $k^0\to 0$ due to the
vanishing quark and gluon masses. Indeed, if all the partonic masses
(virtualities) are neglected, we have in the CMS:
\be t=k^0(-\sqrt{s}  +\sqrt{s} \cos \Theta_2 ) \to 0 \mbox{ at }
k^0\to 0. \ee
 This divergence can be remedied by
introducing a small finite gluon mass $\mu_{cut}$ (cf. the plasmon
mass in~\cite{McLerran:1984ay}).
Indeed, the gluon thermal mass $\mu$ plays the role of a natural
cut-off in the sQGP (cf. section~\ref{sectionoffshellgBr}).

%In Fig.~\ref{sqrt_dependence}, we plot the on-shell cross sections
%introduced above for the dilepton
%production in $q+\bar q\to g+\mu^+\mu^-$
%for $Q=0-4$~GeV, $\mu_{cut}=200$~MeV, $\Lambda=200$~MeV
%and different values of the invariant energy of the $q\bar q$ collision:
%$\sqrt{s}=1.3$~GeV (black solid line),
%$\sqrt{s}=2$~GeV (red dashed line),
%$\sqrt{s}=4$~GeV (blue dotted line),
%$\sqrt{s}=10$~GeV (green dash-dotted line).

%---------------------------------------------------------------------
\subsection{Gluon Compton scattering}

%\begin{figure}
%\begin{center}
%\resizebox{0.46\textwidth}{!}{%
% \includegraphics{GCSon.eps}
%} \caption{Gluon Compton scattering cross sections in the on-shell
%approximation at different $\sqrt{s}$.} \label{GCSon}
%\end{center}
%\end{figure}

In QED, the Compton process refers to elastic scattering of a photon
off a charged object, and has proven to be very important as it
provided early evidence that the electromagnetic wave is
quantized~\cite{Compton:1923zz,Compton:1923zza}. In QCD, the
corresponding process is the {\em gluon Compton scattering}
$g+q(\bar q)\to q(\bar q)+\gamma^*$. The cross section in leading
twist of pQCD~\cite{Altarelli:1977kt} is given by:
\bea \label{Altarelli.GCS} \hspace{-0.4cm} \left( \frac{d^2
\hat\sigma (g+q) }{ d Q^2 d \cos \Theta} \right)^{\mbox{\scriptsize
GCS}} _{\mbox{\scriptsize on-shell}} \hspace{-0.3cm} & = &
\frac{\alpha ^2 e_q^2 \alpha_S}{18 Q^2} \frac{s-Q^2}{s^2
(1+\cos\Theta)} \times \nn & & \left\{
\frac{2s}{s-Q^2}+\frac{s-Q^2}{2s}(1+\cos\Theta)^2 \right. \nn & & \
\  \left. -\frac{2Q^2}{s}(1-\cos\Theta) \right\}. \eea
In terms of Mandelstam variables~\cite{Halzen:1978et} it reads:
 \bea \left( \frac{d \hat \sigma (g+q)}{d Q^2 d t}
\right)^{\mbox{\scriptsize GCS}} _{\mbox{\scriptsize on-shell}}
\hspace{-0.5cm} & = & \frac{e_q^2 \alpha^2 \alpha_S }{9 Q^2}
\frac{s^2+t^2+2 Q^2 u}{- s^3 t} \nn & & \hspace{0.5cm} \times \delta
\left( s+t+u-Q^2 \right), \eea
which is obviously related by crossing transformation to
(\ref{gBrems.stu2}).

\vspace{10pt}
%%%%%%%%%%%%%%%%%%%%%%%%%%%%%%%%%%%%%%%%%%%%%%%%%%%%%%%%%%%%%%%%%%%%%%%%%%%%%%%%%%
\section{dileptons from dynamical quasi-particles}
\label{offshellsection}

Let us now proceed to the calculation of the dilepton production by
effective strongly interacting partonic quasiparticles with broad
spectral functions. Dilepton radiation by the dynamical
quasiparticles proceeds via the elementary processes illustrated in
Figs.~\ref{diagrams} and 2: the basic Drell-Yan $q+\bar q$
annihilation mechanism, Gluon Compton scattering ($q+g\to
\gamma^*+q$ and $\bar q+g\to \gamma^*+\bar q$), quark + anti-quark
annihilation with gluon bremsstrahlung in the final state ($q+\bar
q\to g+\gamma^*$) and gluon decay ($g\to q+\bar q+l^++l^-$).

%---------------------------------------------------------------------
\vspace{10pt}
\subsection{Off-shell $q+\bar q$ in the Drell-Yan mechanism}

\begin{figure*}
\centering \subfigure[ Cross sections]{
\resizebox{0.44\textwidth}{!}{%
 \includegraphics{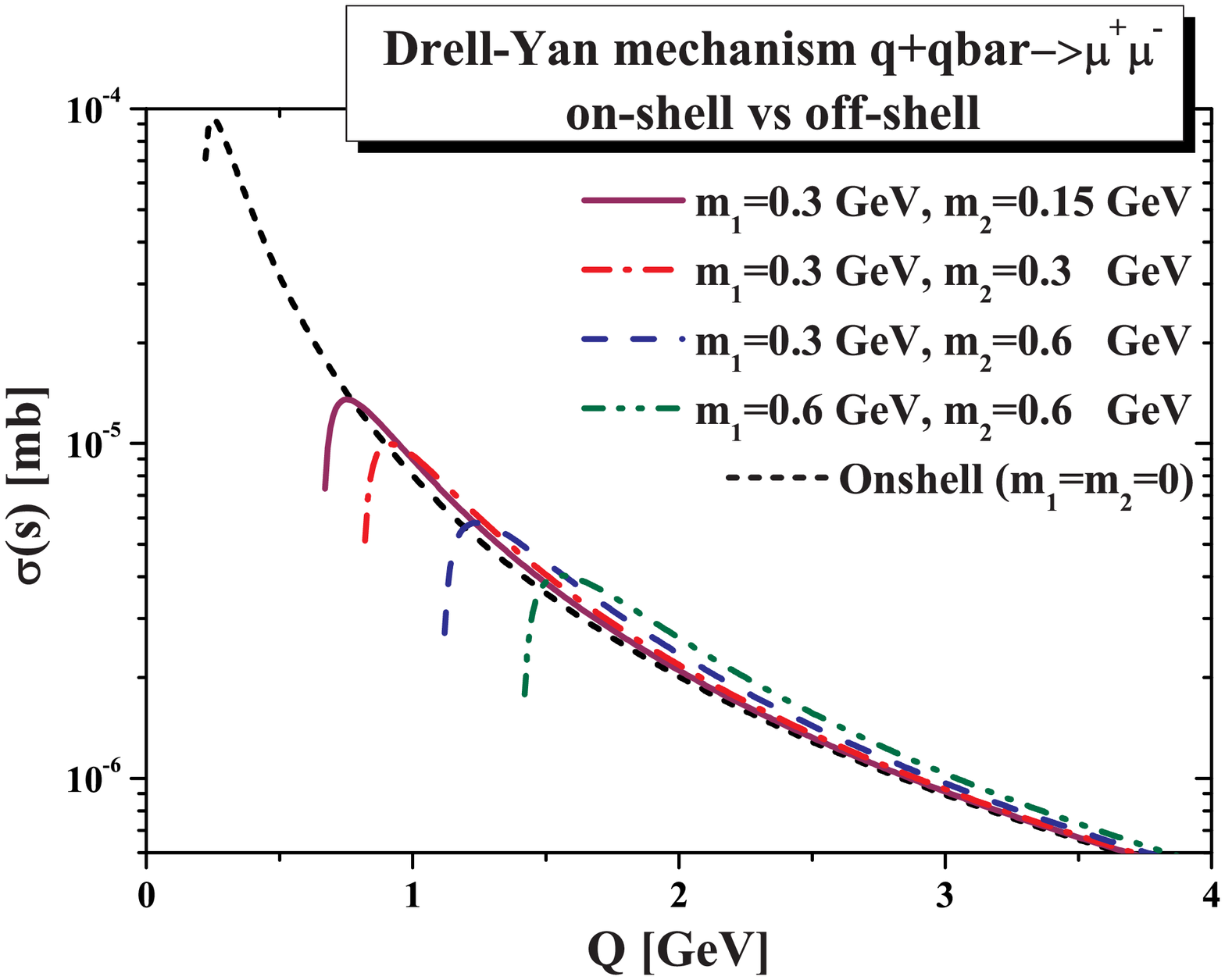}
} } \subfigure[ Ratios of the cross sections]{
\resizebox{0.48\textwidth}{!}{%
 \includegraphics{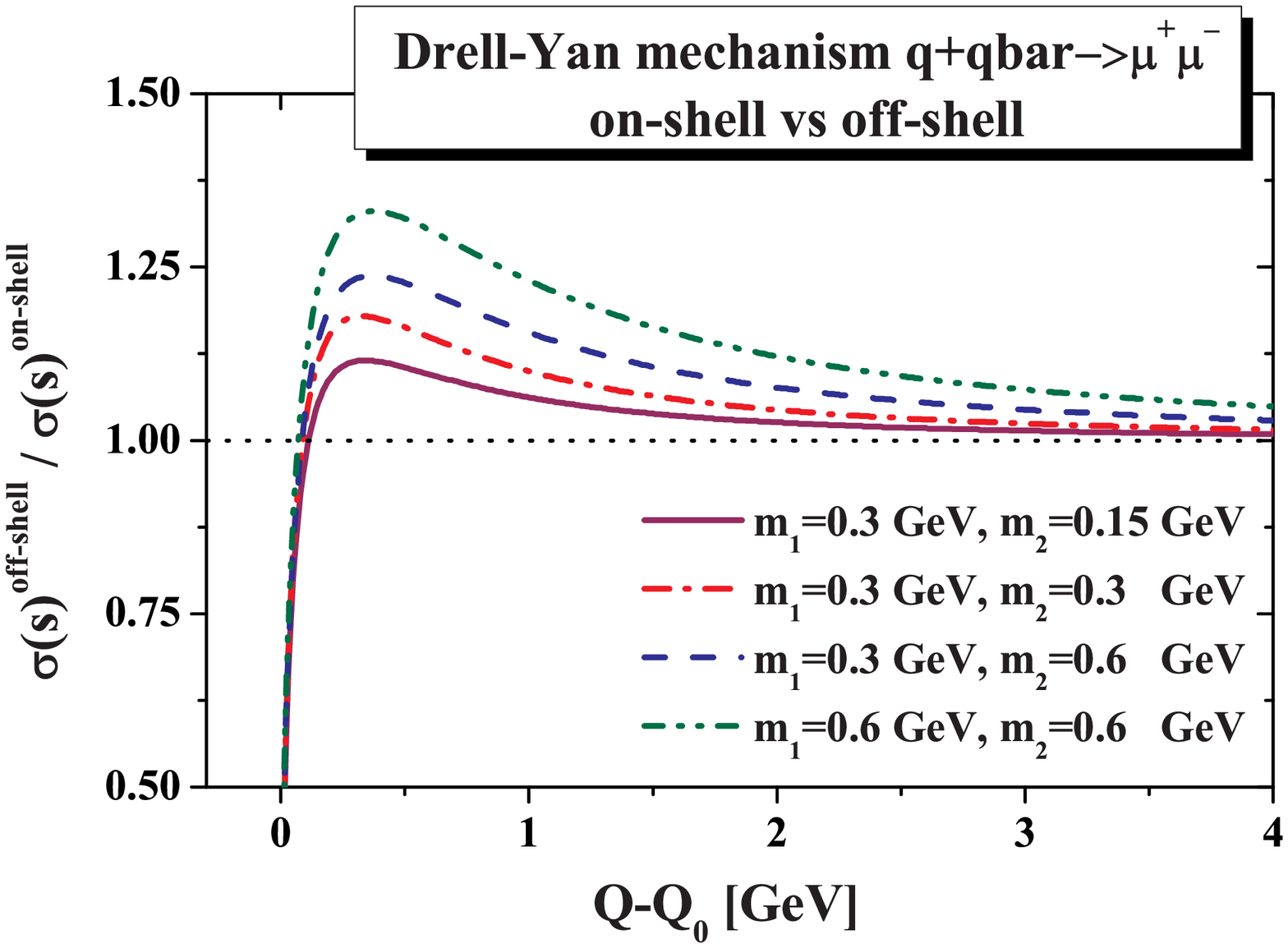}
} }\caption{(color online) Dimuon production cross sections in the
Drell-Yan channel ($q+\bar q\to \mu^+ + \mu^-$). {\bf L.h.s.} The
cross section is presented versus the mass of the muon pair $Q$. The
short dashes (black) line shows the on-shell, i.e. the standard
perturbative result. The other lines show the off-shell cross
section, in which the annihilating quark and antiquark have finite
masses $m_1$ and $m_2$ with different values: $m_1=0.3$~GeV,
$m_2=0.15$~GeV (solid magenta line), $m_1=0.3$~GeV, $m_2=0.3$~GeV
(dash-dotted red line), $m_1=0.3$~GeV, $m_2=0.6$~GeV (dashed blue
line), $m_1=0.6$~GeV, $m_2=0.6$~GeV (dash-dot-dot green line). {\bf
R.h.s.} The ratio of the off-shell cross section to the on-shell
result  for the different values of quark and antiquark masses is
plotted versus $Q-Q_0$, where $Q_0$ is the threshold value for the
lepton pair mass. Line coding as in the figure on the l.h.s.}
\label{DYoffVsOn}
\end{figure*}

For the diagram $(a)$ in Fig.~\ref{diagrams} the off-shell cross
section was obtained in~\cite{Linnyk:2006mv} as:
%
%\begin{widetext}
\bea \label{OffshellSigmaLO} & & \left(
\frac{d^3\hat\sigma(m_1,m_2,\vec{p}_1,\vec{p}_2) }{dQ^2dx_Fdq_T^2}
\right)^{DY} _{off-shell}=
\nn & & \hspace{0.5cm} \kappa' \left[ 2 Q^4 - Q^2 \left( m_1^2 - 6
m_1 m_2 + m_2^2 \right) - \left( m_1^2-m_2^2 \right)^2 \right] \nn
&& \hspace{0.5cm} \times \delta \left( Q^2
        -m_1^2-m_2^2 -
        2 (p_1 \cdot p_2)
        %k_1^-k_2^+
        %- \frac{\left( m_1^2 +\vec{k}_{1\perp}^2 \right)
        %        \left( m_2^2 +\vec{k}_{2\perp}^2 \right)}{k_1^-k_2^+}
        %+ 2 \vec{k}_{1\perp} \cdot \vec{k}_{2\perp}
\right) \nn
&& \hspace{0.5cm} \times \delta\left( x_F - \frac{\sqrt{s}}{s-Q^2}
\left( p_{2 z}-p_{1 z}
%k_2^+ -
%k_1^- + \frac{\left( m_1^2 +\vec{k}_{1\perp} ^2 \right)}{k_1^-} -
%\frac{\left( m_2^2 +\vec{k}_{2\perp} ^2 \right)}{k_2^+}
\right) \right) \nn
&& \hspace{0.5cm} \times \delta \left( q_T^2 - \left(
\vec{p}_{1\perp}+\vec{p}_{2\perp} \right)^2 \right), \eea
%\end{widetext}
%
with
\be
\label{DY5} \kappa' = \frac{2 \pi \alpha^2 e_q^2 %E
}{3 Q^4 N_c 8 \sqrt{(p_1\cdot p_2)^2-m_1^2 m_2^2}}. \ee
In (\ref{OffshellSigmaLO}), $N_c$ is the number of colors,
$e_q$($-e_q$)  is the fractional charge of the quark (antiquark),
$m_i^2$ are the virtualities of the annihilating quark and
antiquark, $p_i$ are their 4-momenta.

Just as in the standard pQCD `on-shell' case, the mass of the
produced dilepton pair is fixed to the invariant energy of the
quark-antiquark collision:
\be
s=Q^2.
\ee
However, the kinematical limit for the minimal dilepton mass
is now higher than in the on-shell case:
\be
Q_0^2\equiv s_0=(m_1+m_2+2 m_{lept})^2 > 4 m_{lept}^2,
\ee
where $m_{lept}$ is the mass of an electron or muon.
Also, the incident current also changes in the off-shell case:
$$
J=\frac{1}{2}\sqrt{(k_1\!\cdot\! k_2)^2\!-\!m_1^2 m_2^2} =
\frac{1}{2}\sqrt{(s-m_1^2-m_2^2)^2\!-\! 4 m_1^2 m_2^2},
$$
compared to $J=s/2$ in the on-shell approximation.
%
%Note that the cross section (\ref{OffshellSigmaLO}) turns out to be gauge
%invariant, as [will be] shown in the Appendix.

The approximation $m_1=m_2\to0$ in formula (\ref{OffshellSigmaLO})
is equivalent to restricting oneself to the leading term in the
twist expansion, that is, in the case of the unpolarized Drell-Yan
process, an expansion in powers of $1/Q$. One can see that in this
limit and  additionally using the collinear kinematics $\vec p_{1
\perp}=\vec p_{2 \perp}=0$ we recover the standard pQCD result
(\ref{OnshellSigmaLO}).

The off-shell cross sections are compared to the leading twist
results in Fig.~\ref{DYoffVsOn}. Dimuon production cross sections in
the Drell-Yan mechanism is plotted on the l.h.s. of
Fig.~\ref{DYoffVsOn} versus the mass of the muon pair $Q=\sqrt{s}$.
The short dashes (black) line shows the on-shell, i.e. the standard
perturbative result. The other lines show the off-shell cross
section, in which the annihilating quark and antiquark have finite
masses $m_1$ and $m_2$ with different values: $m_1=0.3$~GeV,
$m_2=0.15$~GeV (solid magenta line), $m_1=0.3$~GeV, $m_2=0.3$~GeV
(dash-dotted red line), $m_1=0.3$~GeV, $m_2=0.6$~GeV (dashed blue
line), $m_1=0.6$~GeV, $m_2=0.6$~GeV (dash-dot-dot green line).

The importance of higher twist corrections in the Drell-Yan process
is illustrated by the ratio of the off-shell and on-shell integrated
cross section $\sigma(Q)$ on the r.h.s. of Fig.~\ref{DYoffVsOn}. The
ratio of the off-shell cross section to the on-shell result  for the
different values of quark and antiquark masses is plotted versus
$Q-Q_0$, where $Q_0$ is the threshold value for the lepton pair
mass. Line coding as in the figure on the previous figure. With
increasing $Q^2$, off-shell cross sections approach the leading
twist -- on-shell -- result.

In Ref.~\cite{Linnyk:2004mt}, a model for implementing the higher
twist corrections to the Drell-Yan process $p+\bar p\to l^++l^-+X$
was formulated making use of the above cross section, a Breit-Wigner
parametrization for the quark spectral function and non-integrated
parton distributions. The effect of quark off-shellness on the
transverse momentum spectrum of the dileptons produced in $p\bar p$
collisions was found to be large at low $q_T$ and $Q^2$. The
calculations were compared to the data on the triple differential
Drell-Yan cross section $d ^3 \sigma / d Q^2 d x_F d q_T $ from
experiment E866~\cite{WEBB:2003BJ} at Fermilab in $p p$ collisions
at 800~GeV incident energy. Both the slope and magnitude of the
$q_T$ distribution of the Drell-Yan pairs were described by the
adjustment of the spectral function parametrization. The
distribution of the transverse momentum of lepton pairs produced in
the Drell-Yan process off {\em nuclei} $p A \to l^+l^-X$ was also
reproduced within this model~\cite{Linnyk:2006mv}. However, the
effect of quark off-shellness on the dilepton emission in {\em
heavy-ion collisions} ($A+A$) has not been studied untill now. This
question can be addresses by convoluting the off-shell cross section
(\ref{OffshellSigmaLO}) with effective spectral functions $A(m_i)$
for quarks in plasma and with a model for the distribution of quarks
in plasma (cf. Sections IV and V).

%---------------------------------------------------------------------
\vspace{10pt}
\subsection{Off-shell gluon Bremsstrahlung $q \bar q\to g l^+l^-$}
\label{sectionoffshellgBr}

In contrast to the off-shell cross section for the $q\bar q\to
l^+l^-$ process, the off-shell cross section for the $q\bar q$
annihilation {\em with gluon Bremsstrahlung} in the final state has
not been calculated elsewhere. Therefore, we will provide here a
short description of its evaluation.

Starting from the formula for the unpolarized cross section \be
d\sigma = \frac{ \bar{ \Sigma | M_{i\to f} |^2 } \varepsilon_1
\varepsilon_2 \Pi \frac{d^3 p_f}{(2 \pi)^3} } {\sqrt{(p_1 p_2)^2 -
m_1 ^2 m_2 ^2}} \ (2 \pi) ^4 \delta (p_1+p_2-\Sigma p_f), \ee
where the incoming quark and antiquark momenta are $p_1$ and $p_2$,
their masses $m_1$ and $m_2$, respectively; $p_f$ are the momenta of
the outgoing particles, i.e. of the electron (muon) and positron
(anti-muon) and gluon, we note that the dilepton production cross
section can be easily obtained from the cross section for the
production of virtual photons as
\be \frac{d \sigma (l^+l^-)}{d Q^2 dt} = \frac{\alpha}{3 \pi Q^2}
\frac{d \sigma (\gamma^*)}{d t} FF(Q^2,Q_0^2) \ee
with \be FF(Q^2,Q_0^2) = \sqrt{1-\frac{Q_0^2}{Q^2}} \left(1 +
\frac{Q_0^2}{2Q^2}\right),\ee
where $Q_0^2= 4 m_{lept} ^2$, $m_{lept}$ is the lepton mass.

Furthermore, we define the momenta of the internal quark exchanged
in the two relevant diagrams (see Fig.~1) as $p_3 \equiv p_1 - q$,
$\bar p_3 \equiv p_1 - p_2 - p_3 $ and its mass as $m_3$. The final
gluon momentum is $k$ and its mass is $\mu$. Then the matrix element
of the process $q +\bar q \to g + \gamma ^*$ is

\be M=M_a+M_b, \ee where
\bea M_a & = &  - e_q e g_s T_{i j}^l \frac{\epsilon _{\nu} (q)
\epsilon_{\sigma l}(k)}{p_3^2-m_3^2}  \nn && \times u_i (p_1, m_1)
\left[ \gamma^{\nu} (\hat p _3 + m_3) \gamma^{\sigma} \right] v_j (
p_2, m_2)
 \eea
 and
 \bea
M_b & = & - e_q e g_s T_{i j}^l \frac{ \epsilon_{\sigma l} (k)
\epsilon_{\nu} (q) }{ \bar {p_3} ^2 -m_3^2 } \nn & & \times
u_i(p_1,m_1) \left[ \gamma^\eta (\hat{\bar{p_3}} +m_3) \gamma^\nu
\right] v_j (p_2,m_2),
 \eea
$e$ is the electron charge; $e_q$ is the quark fractional charge;
$T_{i j}^l$ is the generator of the SU(3) color group (that will
yield the color factor in the cross section); $\epsilon _{\nu} (q)$
is the polarization vector for the virtual photon with momentum $q$;
$\epsilon_{\sigma l} (k)$ is the polarization vector for the gluon
of momentum $k$ and color $l$; $u_i (p, m)$ is a Dirac spinor for
the quark with momentum $p$, mass $m$ and color $i$; and $v_i (p,m)$
is the  spinor for the anti-quark.

\begin{widetext}

The squared -- and summed over spin polarizations as well as over
color degrees of freedom -- matrix element can be decomposed in the
following summands:
 \be \sum |M|^2 =  \sum M_a^*M_a + \sum
M_b^* M_b +\sum M_a^* M_b + \sum M_b^*M_b, \ee
where the star denotes the complex conjugation.

The spinors for quark states with mass $m_i$ contribute to the
expression for the average matrix element only in the combinations
$\sum \bar u (p, m_i) u (p, m_i)=(\hat p + m_i)$ (cf
~\cite{AkhiezerBerestetsky}) and the correlation functions between
the states with different masses does not enter $|M|^2$. Thus we
find:
\bea \sum M_a ^* M_b & = & - \frac{e_q^2 e^2 g_s^2 \mbox{Tr}\{T^2\}
}{(p_3^2-m_3^2)(\bar{p_3}^2 -m_3^2)} \left[ \mbox{Tr}\left\{
(\hat{p_2}-m_2) \gamma_\sigma (\hat{p_3} + m_3) \gamma_\nu
(\hat{p_1}+m_1) \gamma^\sigma (\hat{\bar{p_3}}+m_3) \gamma^\nu
\right\} \right. \nn & & \left . \phantom{- \frac{e_q^2 e^2 g_s^2
\mbox{Tr}\{T^2\}
}{(p_3^2-m_3^2)(\bar{p_3}^2 -m_3^2)}} %
- \frac{1}{Q^2} \mbox{Tr}\left\{ (\hat{p_2} - m_2) \gamma _\sigma
(\hat{p_3}+m_3) \hat{q} (\hat{p_1}+m_1) \gamma^\sigma
(\hat{\bar{p_3}}+m_3) \hat{q} \right\} \right. \nn & & \left .
\phantom{- \frac{e_q^2 e^2 g_s^2 \mbox{Tr}\{T^2\}
}{(p_3^2-m_3^2)(\bar{p_3}^2 -m_3^2)}} %
-\frac{A}{k^2} \mbox{Tr}\left\{ (\hat{p_2} - m_2) \hat{k}
(\hat{p_3}+m_3) \gamma_\nu (\hat{p_1}+m_1) \hat{k}
(\hat{\bar{p_3}}+m_3) \gamma^\nu \right\} \right. \nn & & \left .
\phantom{- \frac{e_q^2 e^2 g_s^2 \mbox{Tr}\{T^2\}
}{(p_3^2-m_3^2)(\bar{p_3}^2 -m_3^2)}} %
+\frac{A}{k^2 Q^2} \mbox{Tr}\left\{ (\hat{p_2} - m_2) \hat{k}
(\hat{p_3}+m_3) \hat{q} (\hat{p_1}+m_1) \hat{k}
(\hat{\bar{p_3}}+m_3) \hat{q} \right\}
 \right]. \label{M1} \eea
 \bea \sum |M_a|^2 & = & - \frac{e_q^2 e^2 g_s^2 \mbox{Tr}\{T^2\}
}{(p_3^2-m_3^2)^2} \left[ \mbox{Tr}\left\{ \gamma_\sigma (\hat{p_3}
+ m_3) \gamma_\nu (\hat{p_1}+m_1) \gamma^\nu (\hat{p_3} + m_3)
\gamma^\sigma  (\hat{p_2}-m_2) \right\} \right. \nn & & \left .
\phantom{ - \frac{e_q^2 e^2 g_s^2 \mbox{Tr}\{T^2\}
}{(p_3^2-m_3^2)^2}} %
- \frac{1}{Q^2} \mbox{Tr}\left\{  \gamma_\sigma (\hat{p_3} + m_3)
\hat{q}(\hat{p_1}+m_1) \hat{q} (\hat{p_3} + m_3) \gamma^\sigma
(\hat{p_2}-m_2) \right\} \right. \nn & & \left . \phantom{-
\frac{e_q^2 e^2 g_s^2 \mbox{Tr}\{T^2\}
}{(p_3^2-m_3^2)^2}} %
-\frac{A}{k^2} \mbox{Tr}\left\{ \hat{k} (\hat{p_3} + m_3) \gamma_\nu
(\hat{p_1}+m_1) \gamma^\nu (\hat{p_3} + m_3) \hat{k} (\hat{p_2}-m_2)
\right\} \right. \nn & & \left . \phantom{- \frac{e_q^2 e^2 g_s^2
\mbox{Tr}\{T^2\}
}{(p_3^2-m_3^2)^2}} %
+\frac{A}{k^2 Q^2} \mbox{Tr}\left\{ \hat{k} (\hat{p_3} + m_3)
\hat{q} (\hat{p_1}+m_1) \hat{q} (\hat{p_3} + m_3) \hat{k}
(\hat{p_2}-m_2) \right\}
 \right]. \label{M2} \eea
%\end{widetext}

Note that by the transformation $\{ p_3 \to \bar{p_3}, p_1\to p_2,
p_2 \to p_1, m_1 \to -m_2, m_2 \to -m_1 \}$ we readily obtain $\sum
M_b ^* M_a$ from $\sum M_a ^* M_b$ and $\sum |M_b|^2 $ from $\sum
|M_a|^2 $.
 In equations (\ref{M1}) and (\ref{M2}), $A$ sets the
gauge.
%
%For instance, in the generalized renormalizable gauge $A=(1-\lambda)
%k^2 / (k^2-\lambda \mu^2)$, in Feynman gauge $\lambda=0$.
%
We used the feynpar.m~\cite{feynpar.m} package of the Mathematica
program~\cite{Mathematica} to evaluate the traces of the products of
the gamma matrices.
The resulting cross section is (here shown at $A=1$):
\bea
& & \left( \frac{d \hat \sigma(q \, \bar q\to g \, l^+l^-)}{dQ^2 dt}
\right)_{ \! offshell} ^{ \! gBr}  =
 \! - \frac{{\alpha}^2\,\alpha_S\,{e_q}^2\, \,\Theta\left(\phantom{\rule{0pt}{12pt}}\!-Q^2 + s + t - {\mu }^2\right)
\delta (s+t+u-m_1^2-m_2^2-Q^2-\mu^2)
 }{27\,Q^4\,
    {\sqrt{-2 \left( {m_1}^2\,{m_2}^2 \right)  + {\left( {m_1}^2 + {m_2}^2 - s \right) }^2}}\,
    s\,\left( {m_3}^2 - t \right) \,\left( {m_3}^2 - u \right) \,{\mu
    }^2} \phantom{lkhljkhjlkjlkjlkjlkjlkj}
    \nn
& &
\times \left[ {m_3}^6\,\left( s\,t\,\left( s + t \right)  + {\left(
s + 2\,t \right) }^2\,u + \left( s + 4\,t \right) \,u^2 -
Q^4\,\left( t + u \right)  +
     Q^2\,\left( t^2 + u^2 \right)  \right)
  \right.  \nn
& & \left. \vspace{0.4cm}
     + t\,u\,\left( -2\,Q^6\,\left( t + u \right)  +
     2\,t\,u\,\left( 2\,s\,\left( s + t \right)  + \left( 2\,s + t \right) \,u \right)  + 4\,Q^4\,\left( t^2 + u^2 + s\,\left( t + u \right)  \right)
  \right.  \right. \nn
& & \left. \left. \vspace{0.4cm} \vspace{0.4cm}
     -
     Q^2\,\left( 2\,s^2\,\left( t + u \right)  + 4\,s\,\left( t^2 + u^2 \right)  + \left( t + u \right) \,\left( 2\,t^2 - 3\,t\,u + 2\,u^2 \right)  \right)  \right)
   + {m_3}^2\,\left( 2\,Q^6\,{\left( t + u \right) }^2
  \right.  \right. \nn
& & \left. \left. \vspace{0.4cm} \vspace{0.4cm}
   -
     Q^4\,\left( t + u \right) \,\left( 4\,t\,\left( s + t \right)  + \left( 4\,s + t \right) \,u + 4\,u^2 \right)  +
     2\,Q^2\,\left( t^4 + t^3\,u - t^2\,u^2 + t\,u^3 + u^4 + s^2\,{\left( t + u \right) }^2
  \right.  \right.  \right.  \nn
& & \left. \left. \left. \vspace{0.4cm} \vspace{0.4cm}
\vspace{0.4cm}
     + 2\,s\,\left( t + u \right) \,\left( t^2 + u^2 \right)  \right)  +
     t\,u\,\left( -3\,s^2\,\left( t + u \right)  - 2\,s\,\left( t^2 + 3\,t\,u + u^2 \right)  + 2\,\left( t^3 + u^3 \right)  \right)  \right)  -
  {m_3}^4\,\left( -\left( s^2\,{\left( t - u \right) }^2 \right)
  \right.  \right.   \nn
& & \left. \left. \vspace{0.4cm} \vspace{0.4cm}
   + 2\,Q^6\,\left( t + u \right)  -
     2\,s\,\left( t + u \right) \,\left( t^2 - 3\,t\,u + u^2 \right)  + t\,u\,\left( 3\,t^2 + 8\,t\,u + 3\,u^2 \right)  -
     Q^4\,\left( 5\,t^2 + 2\,t\,u + 5\,u^2 + 4\,s\,\left( t + u \right)  \right)
  \right.  \right.   \nn
& & \left. \left. \vspace{0.4cm} \vspace{0.4cm}
     +
     Q^2\,\left( 2\,s^2\,\left( t + u \right)  + 4\,s\,\left( t^2 + u^2 \right)  + 3\,\left( t^3 + u^3 \right)  \right)  \right)  -
  m_1\,m_3\,\left( t\,u\,\left( 2\,Q^4\,\left( t + u \right)  + s^2\,\left( t + u \right)
  \right.  \right. \right.   \nn
& & \left.  \left.  \left. \vspace{0.4cm}  \vspace{0.4cm}
\vspace{0.4cm}
  -
        Q^2\,\left( t^2 + 12\,t\,u + u^2 + 3\,s\,\left( t + u \right)  \right)  \right)  +
     {m_3}^2\,\left( t + u \right) \,\left( -2\,Q^4\,\left( t + u \right)  - s^2\,\left( t + u \right)  +
        Q^2\,\left( t^2 + 12\,t\,u + u^2
  \right.  \right. \right. \right.   \nn
& & \left.  \left.  \left.  \left. \vspace{0.4cm} \vspace{0.4cm}
\vspace{0.4cm} \vspace{0.4cm}
        + 3\,s\,\left( t + u \right)  \right)  \right)  +
     {m_3}^4\,\left( 2\,Q^4\,\left( t + u \right)  + s^2\,\left( t + u \right)  -
        Q^2\,\left( 3\,s\,\left( t + u \right)  + 7\,\left( t^2 + u^2 \right)  \right)  \right)  \right)
  \right.    \nn
& & \left. \vspace{0.4cm}
         +
  \left( -\left( {m_3}^6\,\left( 6\,Q^4 + 2\,s^2 + t^2 + u^2 - 5\,Q^2\,\left( t + u \right)  + 4\,s\,\left( t + u \right)  \right)  \right)
 \right. \right.
\nn & & \left.  \left.  \vspace{0.4cm} \vspace{0.4cm}
  +
     t\,u\,\left( -8\,Q^6 + t\,u\,\left( 2\,s + t + u \right)  + Q^4\,\left( -4\,s + 7\,\left( t + u \right)  \right)  +
        Q^2\,\left( 4\,s^2 + 2\,{\left( t - u \right) }^2 + 5\,s\,\left( t + u \right)  \right)  \right)
 \right. \right.  \nn
& & \left.  \left.  \vspace{0.4cm} \vspace{0.4cm}
        +
     {m_3}^4\,\left( -8\,Q^6 + 2\,t\,{\left( s + t \right) }^2 + \left( 2\,s^2 + 10\,s\,t + t^2 \right) \,u + \left( 4\,s + t \right) \,u^2 + 2\,u^3 +
        Q^4\,\left( -4\,s + 13\,\left( t + u \right)  \right)
 \right. \right. \right.  \nn
& & \left.  \left. \left.  \vspace{0.4cm} \vspace{0.4cm}
\vspace{0.4cm}
        + Q^2\,\left( 4\,s^2 + 5\,s\,\left( t + u \right)  - 2\,\left( t^2 + 8\,t\,u + u^2 \right)  \right)
        \right)  + {m_3}^2\,\left( 8\,Q^6\,\left( t + u \right)  + Q^4\,\left( -7\,t^2 - 20\,t\,u - 7\,u^2 + 4\,s\,\left( t + u \right)  \right)
 \right. \right. \right. \nn & & \left.  \left. \left.  \vspace{0.4cm} \vspace{0.4cm}
\vspace{0.4cm}
        -
        t\,u\,\left( 2\,s^2 + 6\,s\,\left( t + u \right)  + 3\,\left( t^2 + u^2 \right)  \right)  -
        Q^2\,\left( 4\,s^2\,\left( t + u \right)  + \left( t + u \right) \,\left( 4\,t^2 - 13\,t\,u + 4\,u^2 \right)  +
           s\,\left( 7\,t^2 + 6\,t\,u + 7\,u^2 \right)  \right)  \right)
 \right. \right.  \nn
& & \left.  \left. \vspace{0.4cm} \vspace{0.4cm}
           +
     m_1\,m_3\,\left( -\left( {m_3}^4\,
           \left( 10\,Q^4 - 2\,s^2 + t^2 + u^2 - s\,\left( t + u \right)  + 2\,Q^2\,\left( 2\,s + t + u \right)  \right)  \right)  -
        t\,u\,\left( 10\,Q^4 - 2\,s^2 + t^2 + u^2
 \right. \right. \right. \right. \nn
& & \left.  \left. \left. \left.  \vspace{0.4cm} \vspace{0.4cm}
\vspace{0.4cm} \vspace{0.4cm}
        - s\,\left( t + u \right)  + 2\,Q^2\,\left( 2\,s + t + u \right)  \right)  +
        {m_3}^2\,\left( 10\,Q^4\,\left( t + u \right)  + \left( t + u \right) \,\left( -2\,s^2 + t^2 + u^2 - s\,\left( t + u \right)  \right)
 \right. \right. \right. \right. \nonumber \eea
%%%%%%%%%%%%%%%%%%%
\bea & & \left.  \left. \left. \left.  \vspace{0.4cm} \vspace{0.4cm}
\vspace{0.4cm} \vspace{0.4cm}
        +
           4\,Q^2\,\left( -t^2 + 4\,t\,u - u^2 + s\,\left( t + u \right)  \right)  \right)  \right)  \right) \,{\mu }^2 +
  \left( {m_3}^2 - t \right) \,\left( {m_3}^2 - u \right) \,
   \left( 4\,m_1\,m_3\,Q^2 - 12\,{m_3}^2\,Q^2 - 10\,Q^4
\right. \right. \nn & & \left.  \left. \vspace{0.4cm} \vspace{0.4cm}
   - 4\,m_1\,m_3\,s + 4\,{m_3}^2\,s -
     6\,Q^2\,s + 4\,{m_3}^2\,t + 3\,Q^2\,t - s\,t - t^2 + 4\,{m_3}^2\,u + 3\,Q^2\,u - s\,u - 4\,t\,u - u^2 \right) \,{\mu }^4
\right. \nn & & \left.  \vspace{0.4cm}
     +
  \left( {m_3}^2 - t \right) \,\left( {m_3}^2 - u \right) \,
   \left( 2\,\left( m_1 - m_3 \right) \,m_3 - 2\,Q^2 + t + u \right) \,{\mu }^6 +
  2\,{m_2}^6\,\left( {m_3}^2 - t \right) \,\left( {m_3}^2 - u \right) \,\left( 2\,Q^2 - s + {\mu }^2 \right)
\right. \nn & & \left.  \vspace{0.4cm}
   +
  {m_2}^4\,\left( {m_3}^2 - t \right) \,\left( {m_3}^2 - u \right) \,
   \left( 2\,m_1\,m_3\,Q^2 + 2\,{m_3}^2\,Q^2 + 4\,Q^4 - 6\,Q^2\,s + 2\,s^2 - 5\,Q^2\,t + 2\,s\,t - 5\,Q^2\,u + 2\,s\,u
\right. \right. \nn & & \left.  \left. \vspace{0.4cm} \vspace{0.4cm}
   -
     2\,\left( m_3\,\left( m_1 + m_3 \right)  + t + u \right) \,{\mu }^2 - 4\,{\mu }^4 \right)  -
  2\,{m_2}^5\,\left( {m_3}^2 - t \right) \,\left( {m_3}^2 - u \right) \,
   \left( m_1\,s + m_3\,\left( 3\,Q^2 - 2\,s + 3\,{\mu }^2 \right)  \right)
\right. \nn & & \left. \vspace{0.4cm}
   -
  2\,{m_2}^3\,\left( {m_3}^2 - t \right) \,\left( {m_3}^2 - u \right) \,
   \left( -\left( m_1\,s\,\left( -Q^2 + s + t + u - {\mu }^2 \right)  \right)  +
     m_3\,\left( Q^4
\right. \right.  \right. \nn & & \left.  \left.  \left.
\vspace{0.4cm} \vspace{0.4cm}
 \vspace{0.4cm}
     - 3\,Q^2\,\left( s + t + u \right)  + 2\,s\,\left( s + t + u \right)  - 3\,\left( s + t + u \right) \,{\mu }^2 + {\mu }^4 \right)
     \right)  - {m_2}^2\,\left( {m_3}^2 - t \right) \,\left( {m_3}^2 - u \right) \,
   \left( 2\,s\,t\,u + \left( 4\,s^2
\right. \right.  \right. \nn & & \left.  \left.  \left.
\vspace{0.4cm} \vspace{0.4cm}
 \vspace{0.4cm}
   - 2\,t\,u + 3\,s\,\left( t + u \right)  \right) \,{\mu }^2 - 5\,\left( 2\,s + t + u \right) \,{\mu }^4 + 6\,{\mu }^6 +
     5\,Q^4\,\left( t + u - 2\,{\mu }^2 \right)  - 2\,{m_3}^2\,\left( Q - \mu  \right) \,\left( Q + \mu  \right) \,
      \left( 2\,Q^2
\right. \right.  \right. \nn & & \left.  \left.  \left.
\vspace{0.4cm} \vspace{0.4cm}
 \vspace{0.4cm}
      - 2\,s - t - u + 2\,{\mu }^2 \right)  - Q^2\,\left( 5\,t^2 - 2\,t\,u + 5\,u^2 + 5\,s\,\left( t + u \right)  + 2\,s\,{\mu }^2 -
        8\,\left( t + u \right) \,{\mu }^2 + 2\,{\mu }^4 \right)  +
     2\,m_1\,m_3\,\left( s^2
\right. \right.  \right. \nn & & \left.  \left.  \left.
\vspace{0.4cm} \vspace{0.4cm}
 \vspace{0.4cm}
     - \left( 3\,s + t + u \right) \,{\mu }^2 + 2\,{\mu }^4 + Q^2\,\left( -s + t + u - 4\,{\mu }^2 \right)  \right)
     \right)  - m_2\,\left( m_1\,\left( -2\,{m_3}^6\,
         \left( Q^4 - 2\,Q^2\,\left( s - 5\,{\mu }^2 \right)  + {\left( s - {\mu }^2 \right) }^2 \right)
\right. \right.  \right. \nn & & \left.  \left.  \left.
\vspace{0.4cm} \vspace{0.4cm}
 \vspace{0.4cm}
         +
        t\,u\,\left( 2\,s\,t\,u + \left( {\left( t - u \right) }^2 + s\,\left( t + u \right)  \right) \,{\mu }^2 - \left( t + u \right) \,{\mu }^4 -
           Q^4\,\left( t + u - 4\,{\mu }^2 \right)  + Q^2\,\left( {\left( t - u \right) }^2 + s\,\left( t + u \right)
\right. \right. \right. \right.  \right. \nn & & \left.  \left.
\left. \left. \left. \vspace{0.4cm} \vspace{0.4cm} \vspace{0.4cm}
\vspace{0.4cm}
 \vspace{0.4cm}
 - 26\,\left( t + u \right) \,{\mu }^2 +
              4\,{\mu }^4 \right)  \right)  + {m_3}^4\,\left( 2\,s\,t\,u + 2\,s^2\,\left( t + u \right)  + {\left( t - u \right) }^2\,{\mu }^2 -
           3\,s\,\left( t + u \right) \,{\mu }^2 + \left( t + u \right) \,{\mu }^4
\right. \right. \right. \right.   \nn & & \left.  \left. \left.
\left. \vspace{0.4cm} \vspace{0.4cm} \vspace{0.4cm}
 \vspace{0.4cm}
           + Q^4\,\left( t + u + 4\,{\mu }^2 \right)  +
           Q^2\,\left( {\left( t - u \right) }^2 - 3\,s\,\left( t + u \right)  - 6\,\left( t + u \right) \,{\mu }^2 + 4\,{\mu }^4 \right)  \right)  +
        {m_3}^2\,\left( -2\,s\,t\,u\,\left( s + t + u \right)
\right. \right. \right. \right.   \nn & & \left.  \left. \left.
\left. \vspace{0.4cm} \vspace{0.4cm} \vspace{0.4cm}
 \vspace{0.4cm}
         - {\left( t - u \right) }^2\,\left( s + t + u \right) \,{\mu }^2 +
           \left( t^2 + u^2 \right) \,{\mu }^4 + Q^4\,\left( t^2 + u^2 - 4\,\left( t + u \right) \,{\mu }^2 \right)  +
           Q^2\,\left( -\left( {\left( t - u \right) }^2\,\left( s + t + u \right)  \right)
\right. \right. \right. \right. \right.   \nn & & \left.  \left.
\left. \left. \left. \vspace{0.4cm} \vspace{0.4cm} \vspace{0.4cm}
\vspace{0.4cm}
 \vspace{0.4cm}
           - 2\,\left( 5\,t^2 - 52\,t\,u + 5\,u^2 \right) \,{\mu }^2 -
              4\,\left( t + u \right) \,{\mu }^4 \right)  \right)  \right)  +
     m_3\,\left( t\,u\,\left( -4\,Q^6 - 4\,s\,t\,u + Q^4\,\left( 8\,s + t + u \right)  +
           Q^2\,\left( -4\,s^2 + t^2
\right. \right. \right. \right. \right.   \nn & & \left.  \left.
\left. \left. \left. \vspace{0.4cm} \vspace{0.4cm} \vspace{0.4cm}
\vspace{0.4cm}
 \vspace{0.4cm}
           + 4\,t\,u + u^2 - s\,\left( t + u \right)  \right)  \right)  -
        \left( 12\,t^2\,{\left( s + t \right) }^2 + t\,\left( -4\,Q^4 + 4\,s^2 + 19\,s\,t + 17\,t^2 + Q^2\,\left( -12\,s + 8\,t \right)  \right) \,u +
           \left( 12\,s^2
\right. \right. \right. \right. \right.   \nn & & \left.  \left.
\left. \left. \left. \vspace{0.4cm} \vspace{0.4cm} \vspace{0.4cm}
\vspace{0.4cm}
 \vspace{0.4cm}
           + 8\,Q^2\,t + 19\,s\,t - 4\,t^2 \right) \,u^2 + \left( 24\,s + 17\,t \right) \,u^3 + 12\,u^4 \right) \,{\mu }^2 +
        t\,u\,\left( 4\,Q^2 + 8\,s + t + u \right) \,{\mu }^4 - 4\,t\,u\,{\mu }^6
\right. \right. \right.   \nn & & \left.  \left. \left.
\vspace{0.4cm} \vspace{0.4cm}
 \vspace{0.4cm}
        +
        {m_3}^2\,\left( -\left( \left( t + u \right) \,\left( -4\,Q^6 - 4\,s\,t\,u + Q^4\,\left( 8\,s + t + u \right)  +
                Q^2\,\left( -4\,s^2 + t^2 + 4\,t\,u + u^2 - s\,\left( t + u \right)  \right)  \right)  \right)
\right. \right. \right. \right.   \nn & & \left.  \left. \left.
\left. \vspace{0.4cm} \vspace{0.4cm} \vspace{0.4cm}
 \vspace{0.4cm}
                +
           \left( -4\,Q^4\,\left( t + u \right)  + 28\,s^2\,\left( t + u \right)  + \left( t + u \right) \,\left( 23\,t^2 + 8\,t\,u + 23\,u^2 \right)  +
              s\,\left( 49\,t^2 + 74\,t\,u + 49\,u^2 \right)
\right. \right. \right. \right. \right.   \nn & & \left.  \left.
\left. \left. \left. \vspace{0.4cm} \vspace{0.4cm} \vspace{0.4cm}
\vspace{0.4cm}
 \vspace{0.4cm}
               + 4\,Q^2\,\left( t + u \right) \,\left( -3\,s + 2\,\left( t + u \right)  \right)  \right) \,{\mu }^2 -
           \left( t + u \right) \,\left( 4\,Q^2 + 8\,s + t + u \right) \,{\mu }^4 + 4\,\left( t + u \right) \,{\mu }^6 \right)  +
        {m_3}^4\,\left( -4\,Q^6
\right. \right. \right. \right.   \nn & & \left.  \left. \left.
\left. \vspace{0.4cm} \vspace{0.4cm} \vspace{0.4cm}
 \vspace{0.4cm}
        - 4\,s\,t\,u - \left( 28\,s^2 + 11\,t^2 + 32\,t\,u + 11\,u^2 + 43\,s\,\left( t + u \right)  \right) \,{\mu }^2 +
           \left( 8\,s + t + u \right) \,{\mu }^4 - 4\,{\mu }^6 + Q^4\,\left( 8\,s + t + u + 4\,{\mu }^2 \right)
\right. \right. \right. \right.   \nn & & \left.  \left. \left.
\left. \vspace{0.4cm} \vspace{0.4cm} \vspace{0.4cm} \vspace{0.4cm}
          +
           Q^2\,\left( -4\,s^2 + t^2 + 4\,t\,u + u^2 - 8\,\left( t + u \right) \,{\mu }^2 + 4\,{\mu }^4 - s\,\left( t + u - 12\,{\mu }^2 \right)  \right)  \right)
        \right)  \right) \right]
\label{gBr_offshell}
\eea

\end{widetext}

\begin{figure*}
\centering \subfigure[ Cross sections]{
\resizebox{0.48\textwidth}{!}{%
 \includegraphics{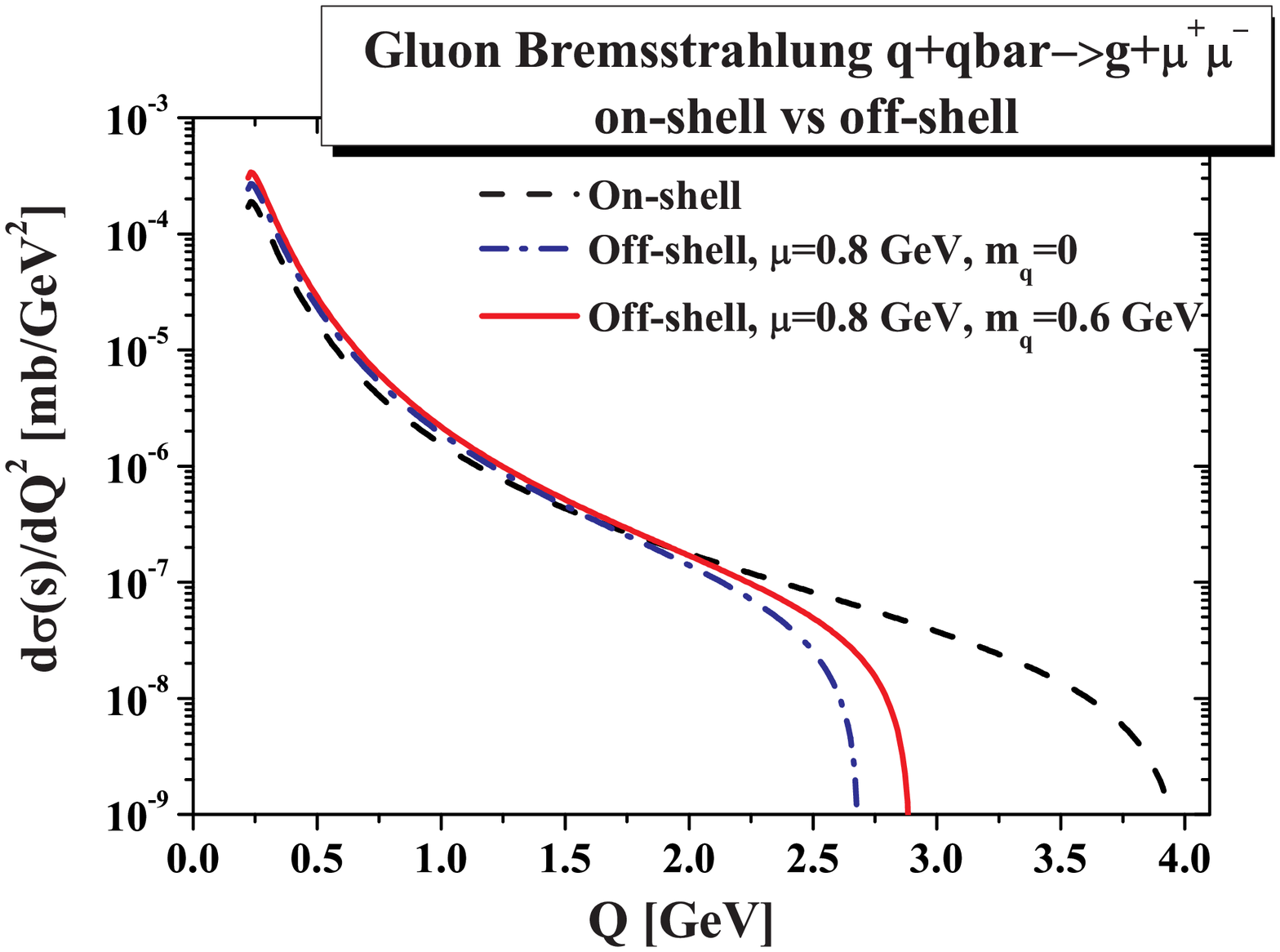}
} } \subfigure[ Ratios of the cross sections]{
\resizebox{0.47\textwidth}{!}{%
 \includegraphics{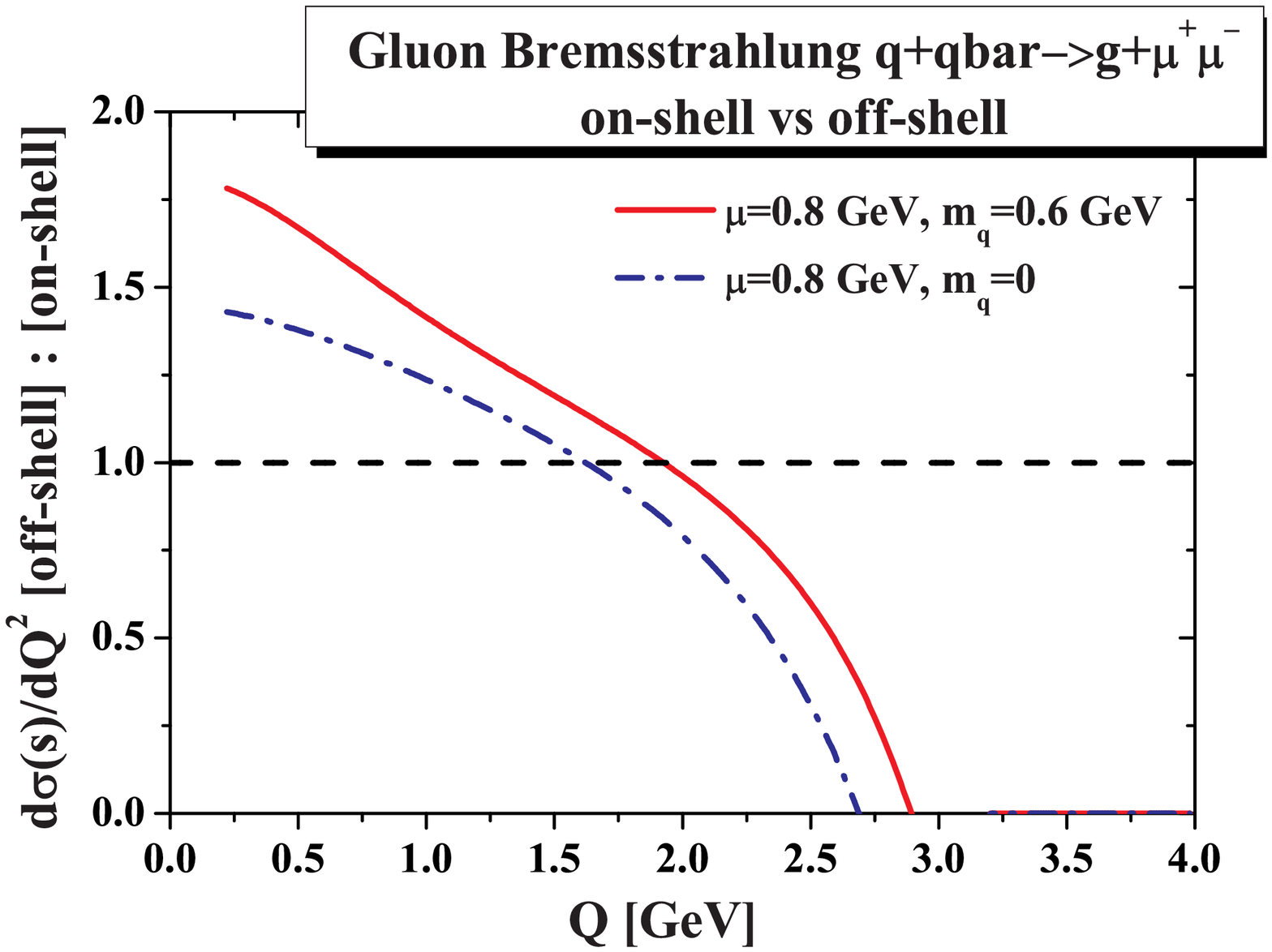}
} }\caption{(color online) Comparison of off-shell and
  on-shell gluon Bremsstrahlung $q+\bar q\to g+\mu^+\mu^-$
  cross sections at $\sqrt{s}=4$~GeV. {\bf L.h.s.} Dashed black line shows
  the on-shell cross section with $\mu_{cut}=0.206$~GeV, blue dash-dotted line
  presents the off-shell cross section for the gluon mass fixed to
  $\mu=0.8$~GeV and on-shell quark and anti-quark ($m_1=m_2=m_3=0$).
  Red solid  line gives the off-shell result for $\mu=0.8$~GeV,
  $m_1=m_2=m_3=m_q=0.6$~GeV. {\bf R.h.s.} The ratio of off-shell to
  on-shell cross sections. Line coding is as on the l.h.s. plot.} \label{gBRoffshell}
\end{figure*}

\begin{figure*}
\centering \subfigure[ Cross sections]{
\resizebox{0.48\textwidth}{!}{%
 \includegraphics{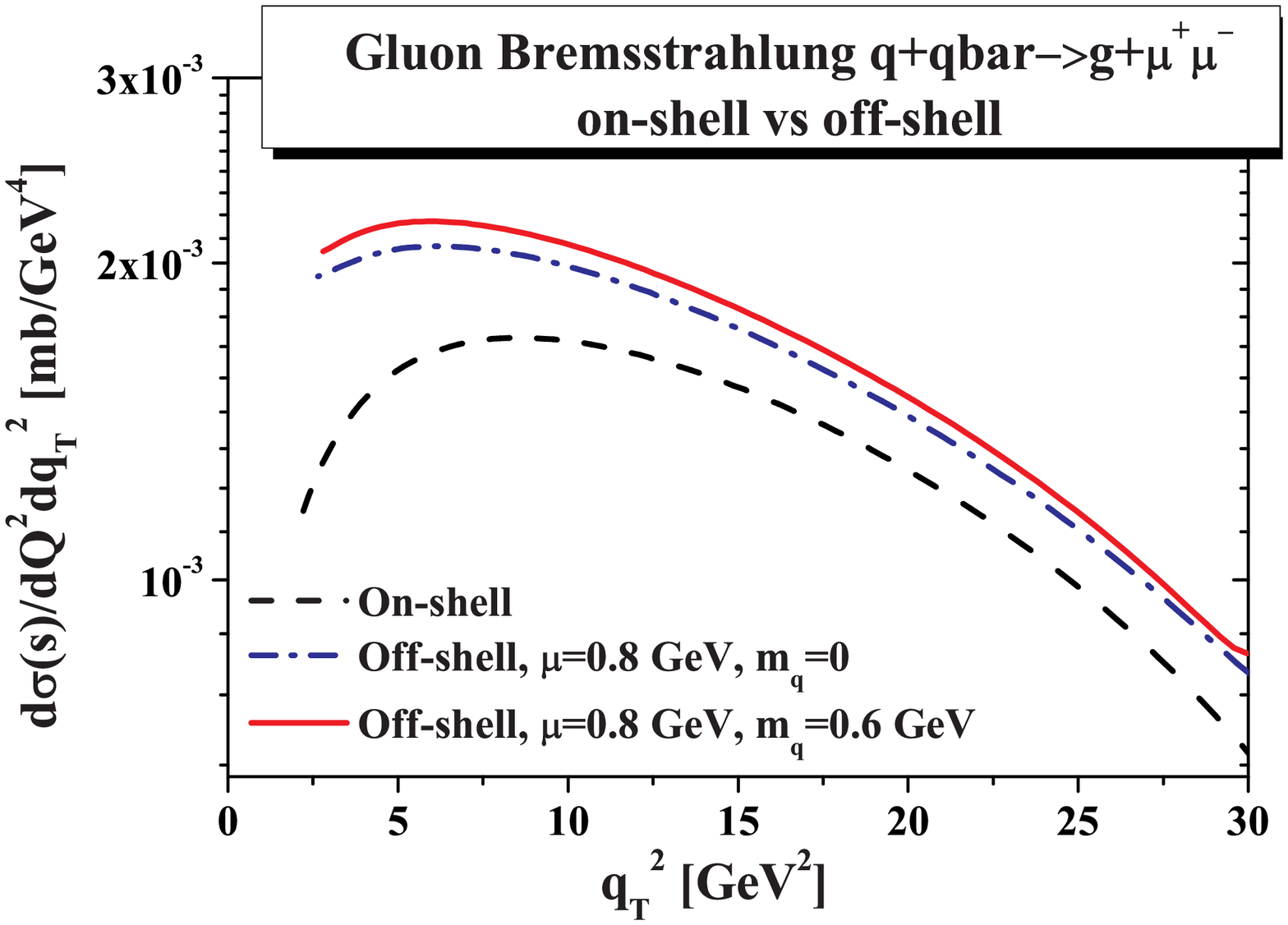}
} } \subfigure[ Ratios of the cross sections]{
\resizebox{0.47\textwidth}{!}{%
 \includegraphics{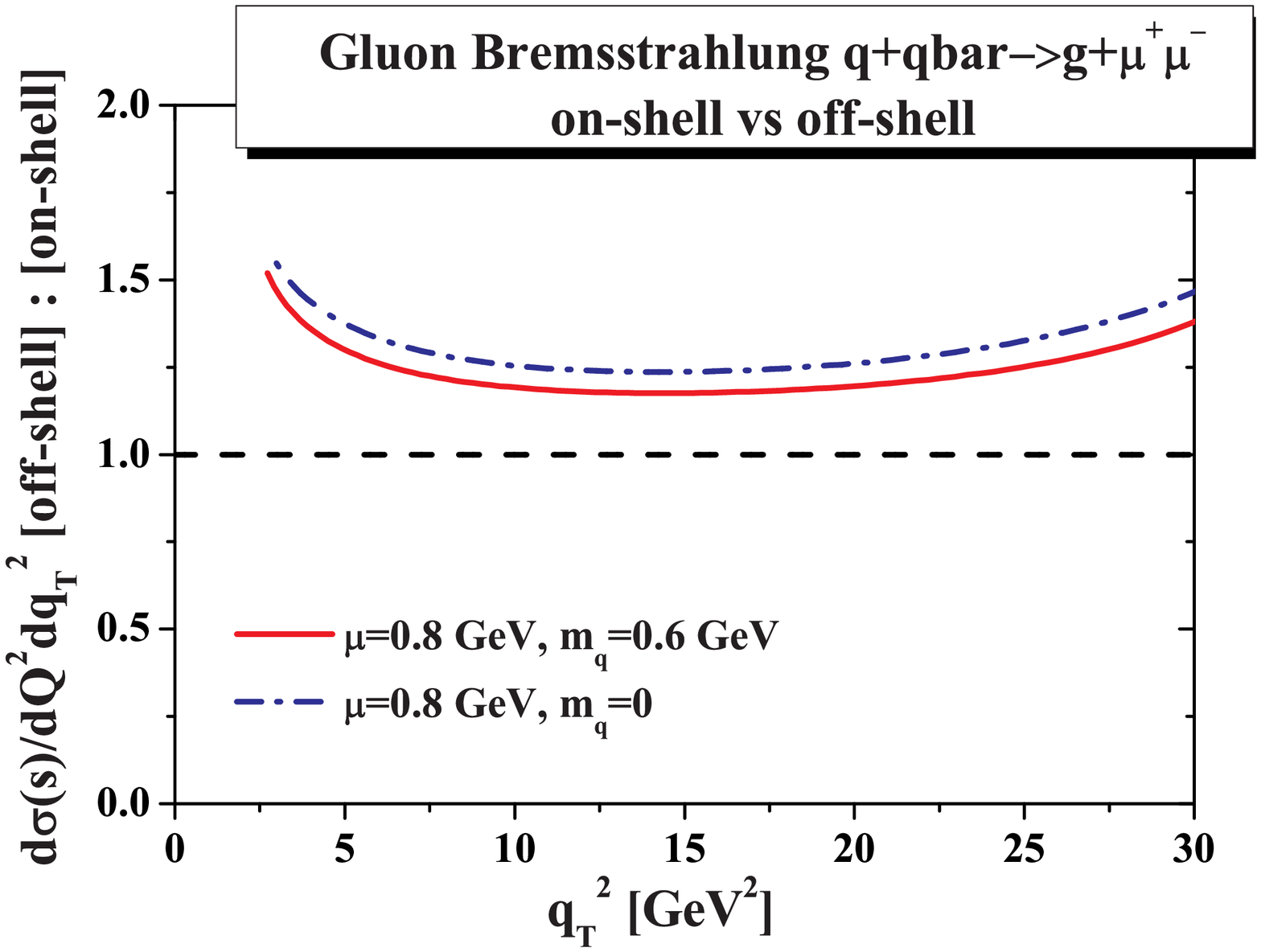}
} }\caption{(color online) Comparison of the transverse momentum
distributions of muon pairs produced in the gluon Bremsstrahlung
$q+\bar q\to g+\mu^+\mu^-$ channel in the off-shell and on-shell
cases. {\bf L.h.s.} Dashed black line shows
  the differential on-shell cross section with $\mu_{cut}=0.206$~GeV, blue dash-dotted line
  presents the off-shell cross section for the gluon mass fixed to
  $\mu=0.8$~GeV and on-shell quark and anti-quark ($m_1=m_2=m_3=0$).
  Red solid line gives the off-shell result for $\mu=0.8$~GeV,
  $m_1=m_2=m_3=m_q=0.6$~GeV. {\bf R.h.s.} The ratio of off-shell to
  on-shell cross sections. Line coding is as on the l.h.s. plot.
   } \label{gBRoffshell2}
\end{figure*}

In the off-shell case, the dilepton mass range is $ 4 m_{lept}^2 \!
< \! Q^2 \! < \! (\sqrt{s}-m_q)^2 $, while the kinematical limits
for the momentum transfer $t$ are
\be \label{eqT1} t_{min}^{max}=-\frac{s}{2} (C_1\pm C_2), \ee
where
\bea C_1 \! & \! = \! & \! 1 - (\beta_!+\beta_2+\beta_3+\beta_4 ) %\nn & &
+ (\beta_1-\beta_2) (\beta_3-\beta_4), \nn C_2 \! & \! = \! & \!
\sqrt{(1-\beta_1-\beta_2)^2-4 \beta_1 \beta_2} \nn & &
\phantom{(1-\beta_3-\beta_4)^2} \times \sqrt{(1-\beta_3-\beta_4)^2 -
4 \beta_3 \beta _4} \label{eqT2} \eea
with
\be \beta_1=m_1^2/s,\mbox{ }\beta_2=m_2^2/s,\mbox{
}\beta_3=Q^2/s,\mbox{ }\beta_4=\mu^2/s.
 \ee
%
%
%Note that in the integration over $t$ from $t_{\mbox{min}}$ to
%$t_{\mbox{max}}$ the cut-off $\Lambda^2$ should be
%incorporated (cf. Section~\ref{cutoffintro}).
Additionally, we note that there is a threshold in the CMS energy
$\sqrt{s}$ for the $q+\bar q$ interaction:
\be s \ge \max\{ (m_1+m_2)^2, (\mu + Q)^2 \}. \ee

One can easily check that the expression (\ref{gBr_offshell}) for
$m_i\to0$ approaches the leading twist pQCD result,
% given by (\ref{gBrems.stuMu}),
where $\mu_{cut}=\mu$. We illustrate this in Fig.~\ref{gBRoffshell}.
In Fig.~\ref{gBRoffshell}, the off-shell cross sections for the
quark annihilation with gluon bremsstrahlung mechanism at various
values of quark and gluon off-shellnesses (masses) are compared to
the on-shell (pQCD) result. Dashed black line shows
  the on-shell cross section with $\mu_{cut}=0.206$~GeV, red solid line
  presents the off-shell cross section for the gluon mass fixed to
  $\mu=0.8$~GeV and on-shell quark and anti-quark ($m_1=m_2=m_3=0$).
  Blue dash-dotted line gives the off-shell result for $\mu=0.8$~GeV,
  $m_1=m_2=m_3=m_q=0.6$~GeV.
One readily notices the shift of the maximum allowed mass of the
pair to a lower value (in order to produce a massive gluon in the
final state). For the rest of the $Q$ values, the effect of the
quark and gluon off-shellness reaches at most 50\%, as is seen in
the ratios of the cross sections, plotted in the r.h.s. of
Fig.~\ref{gBRoffshell}.

Next we compare the double differential off-shell and on-shell cross
sections. The results for the transverse momentum distributions of
the dileptons are presented in Fig.~\ref{gBRoffshell2}. Solid black
line shows
  the differential on-shell cross section with $\mu_{cut}=0.206$~GeV, blue dashed line
  presents the off-shell cross section for the gluon mass fixed to
  $\mu=0.8$~GeV and on-shell quark and anti-quark ($m_1=m_2=m_3=0$).
  Red dash-dotted line gives the off-shell result for $\mu=0.8$~GeV,
  $m_1=m_2=m_3=m_q=0.6$~GeV. Again, we find the largest effect on
  the edge of the phase space, at the minimal $q_T$.

%---------------------------------------------------------------------
\subsection{Off-shell gluon Compton scattering $gq\to q l^+l^-$}

The cross section for the gluon Compton scattering can be calculated
analogously to the calculation of the gluon Bremststrahlung cross
section (\ref{gBr_offshell}) above. On the other hand, the cross
sections for $g+q\to q+l^++l^-$ and $g+\bar q\to \bar q+l^++l^-$ is
readily obtained from (\ref{gBr_offshell}) by the crossing
transformation.

%The expression for this process cross section is more compact, if we
%approximate $m_1=m_2=m_3\equiv m_q$, i.e. that all the quark masses
%are equal:
%%
%\begin{widetext}
%\bea \left( \frac{d \hat \sigma(g+q)}{dQ^2 dt} \right)_{\! \!
%offshell} \! \! \! \! & \! \! = \! \! &  \! \frac{e_q^2 \alpha^2
%\alpha_S }{9Q^2 s} \times \nn && \left\{ \frac{\left( t- \mu^2 - 3
%m_q^2 \right) \left[ 2 Q^2 (Q^2 -2 t+7 m_q^2) + t (2 t -m_q^2) -
%m_q^4 \right]}{\left[ (t-m_q^2-\mu^2)^2 - 4 m_q^2 \mu^2 \right]
%(t-m_q^2)^2} \right. \nn & & \left.  + \frac{\left( s- \mu^2 - 3
%m_q^2 \right) \left[ 2 Q^2 (Q^2 -2 s+7 m_q^2) + s (2 s -m_q^2) -
%m_q^4 \right]}{\left[ (s-m_q^2-\mu^2)^2 - 4 m_q^2 \mu^2 \right]
%(s-m_q^2)^2} \right\},  \eea

Kinematic limits on $s$, $t$, $Q^2$ in the off-shell GCS process are
analogous to the $q+\bar q$ case. In the off-shell case, the
dilepton mass range is $ 4 m_{lept}^2 \! < \! Q^2 \! < \!
(\sqrt{s}-m_q)^2 $, while the kinematical limits on the momentum
transfer $t$ are given by the formulae (\ref{eqT1})-(\ref{eqT2})
with
\be \beta_1=m_1^2/s,\mbox{ }\beta_2=\mu^2/s,\mbox{
}\beta_3=Q^2/s,\mbox{ }\beta_4=m_2^2/s, \ee
%
%
%\bea t_{\mbox{min}}^{\mbox{max}} \! & \! = \! & \! m_q^2 + Q^2 -
%\frac{(s+Q^2-m_q^2)(s+m_q^2-\mu^2)}{2 s} \pm \nn & & \hspace{-2cm}
%\frac{\sqrt{(s-Q^2-m_q^2)^2-4Q^2 m_q^2}
%\sqrt{(s-m_q^2-\mu^2)^2-4m_q^2 \mu^2}}{2s}, \nonumber \eea
%
while $$  s \ge \max\{ (m_1+\mu)^2, (m_2 + Q)^2 \}. \ $$

%---------------------------------------------------------------------
\subsection{Virtual gluon decay $g\to q \bar q l^+l^-$ and virtual quark decay
$q\to g q l^+l^-$}

Although the process of real gluon decay $g\to q+\bar q + l^=l^-$ is
forbidden kinematically for perturbative particles, it has a finite
region of validity, if the gluon is off-shell, due to its broad with
and finite pole mass. Analogously, the virtual quark decay is also
possible in the off-shell case. We present the Feynman diagrams for
the corresponding processes in Fig.~\ref{diagrams2}. The off-shell
cross section for these processes can straightforwardly be obtained
from (\ref{gBr_offshell}) by the crossing relation. For example, the
cross section for $q\to g q l^+ l^-$ is obtained from $q \bar \to g
l^+ l^-$ by changing $p_2 \to -p_2$.

Note that, in order to obtain the dilepton rates, the elementary
cross sections have to be consequently convoluted with the effective
spectral functions for quarks and gluons. The magnitude and shape of
the contributions of the virtual decays to the dilepton rates is
very sensitive to the final choice of the spectral function. In the
DQPM the contribution from gluon decay is higher than that from
virtual quark decay, since the gluonic quasi-particles are more
 massive and broader than the quarks~\cite{Cassing:2007nb},
 and thus the kinematically allowed region is larger for the virtual gluon
 than for the virtual quark decay.

 On the other hand, within the DQPM parametrizations
 for the partonic spectral functions, both processes
 presented by the diagrams in Fig.~\ref{diagrams2} generate little
 dilepton yield anywhere except the extremely low masses: $Q\approx 2 m_{muon}$.
 Therefore, we refrain from plotting here the contributions explicitly and also do
 not consider them in the next section dedicated to the comparison of the yields
 from different mechanisms.

%%%%%%%%%%%%%%%%%%%%%%%%%%%%%%%%%%%%%%%%%%%%%%%%%%%%%%%%%%%%%%%%%%%%%%%%%%%%%%%%%%
\vspace{10pt}
\section{contributions of the different processes to the dilepton rates from QGP}
\label{relative}

\begin{figure}
\begin{center}
        \resizebox{0.46\textwidth}{!}{%
        \includegraphics{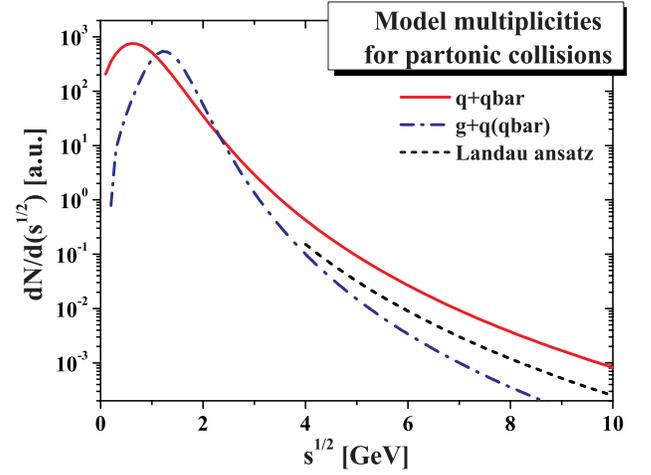}
        }
\caption{(color online) Multiplicities of $q+\bar q$ and
$g+q$($g+\bar q$) collisions. The red solid line represents the
multiplicities of the $q+\bar q$ collisions and the blue dashed line
represents the probability for the
 gluon+quark(antiquark) interaction. The black dashed line shows for comparison
 the prediction in the Landau model for the heavy-ion collision -- the power law
 fall as $(\sqrt{s})^{-7}$.
}
\label{ModelDNDS}
\end{center}
\end{figure}

In the following we are interested in the relative importance of the
different processes and their contributions to the dilepton yield of
the strongly coupled quark-gluon plasma.

\begin{figure}
        \resizebox{0.46\textwidth}{!}{%
        \includegraphics{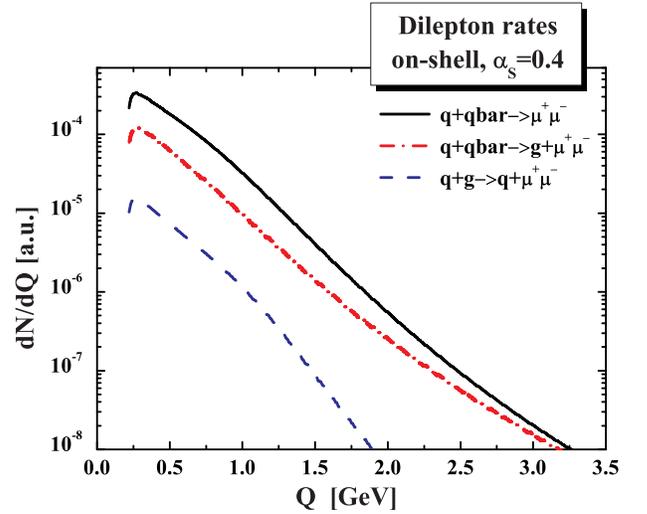}
        }
\caption{(color online) Dimuon rates $d N/dQ$ from QGP calculated
using the cross sections in the on-shell approximation,
$\alpha_S=0.4$. Black solid line shows the contribution of the
Drell-Yan channel ($q+\bar q\to \mu^+\mu^-$), red dash-dotted line
represents the contribution of the channel $q+\bar q\to
g+\mu^-\mu^+$, blue dashed line shows the contribution of the
channel $q+g\to q+\mu^+\mu^-$. The rates are in `arbitrary units',
reflecting our use of simplistic quark and gluon distributions in
the QGP.} \label{Spectra_with_constant alpha}
\end{figure}

Due to the factorization property proven in~\cite{McLerran:1984ay},
the dilepton emission from the QGP created
in the heavy-ion collision is given by the convolution of the
elementary sub-process cross sections (describing quark/gluon
interactions resulting in the emission of dileptons) with the
structure functions that characterize the properties and evolution
of the plasma (encoded in the distribution of the quarks and gluons
with different momenta and virtualities):
\bea  \frac{d ^3 \sigma ^{\mbox{\small QGP}} }{dQ^2dx_Fdq_T^2} & = &
\nn
&& \hspace{-1.5cm} \sum _{abc} \! \int \! d \hat s \! \int_0^\infty
\! \! d m^{i}_1 \! \int_0^\infty \! \! d m^{i}_2 \! \int_0^\infty \!\!
d \mu ^f \, F_{ab}(\hat s,m^{i}_1,m^{i}_2)\nn
&& \hspace{-0.8cm} \times  \mbox{A}_c (\mu^f) \frac{d^3 \hat \sigma
_{abc} (\hat s,m^{i}_1,m^{i}_2,\mu^{f}) }{dQ^2dx_Fdq_T^2}, \
\label{factorization} \eea
where $m^{i}_1$ and $m^{i}_2$ are the off-shellnesses (i.e.
virtualities) of the incoming partons, $\mu^f$ in the off-shellness
of the outgoing parton, the indices $a,b,c$ denote quark, antiquark
or gluon so that all the considered mechanisms are covered. The
cross sections $\hat \sigma _{abc} (\hat s,m^{i1},m^{i2},\mu^{f})$
for the different processes have been derived in the previous
Section.

\begin{figure}
\begin{center}
        \resizebox{0.5\textwidth}{!}{%
        \includegraphics{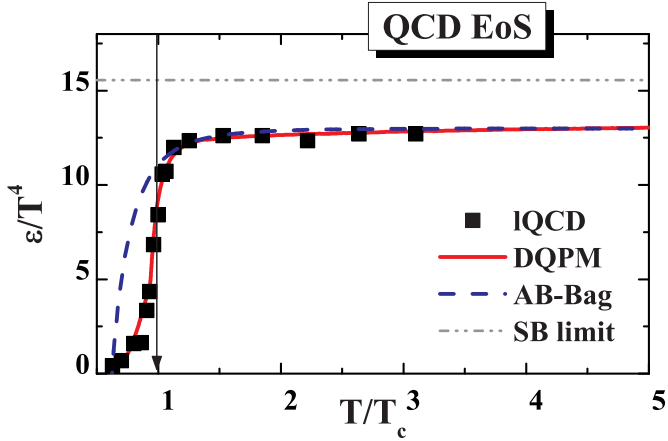}
        }
\caption{(color online) QCD energy density versus temperature from
lattice QCD (square symbols)~\protect{\cite{Karsch:2000ps}}, the
 DQPM approach (red solid line)~\protect{\cite{Cassing:2007nb}} and
 the
 `AB-Bag' model (blue dashed line)~\protect{\cite{Begun:2010md}}.
 The grey dash-dotted
 line shows the Stefan-Boltzmann limit.
 The arrow indicates the critical temperature.}
\label{EoS}
\end{center}
\end{figure}

In (\ref{factorization}), we integrate over the motion of partons,
but also over their virtualities by employing phenomenological
structure functions $F_{ab}$ that depend on the invariant energy
$\hat s$ of the partonic sub-process as well as on the virtualities
of the incoming partons and the spectral function $\mbox{A}(\mu^f)$
for the virtuality distribution of the parton in the final state.
Here should in principle be a two particle correlator, but we work
in the 2PI approximation so that the parton in the sQGP can be
characterized by a single-particle distribution, therefore we assume
that the plasma structure function can be approximated by
\be F_{ab}(\hat s,m_{1},m_{2}) = A_a (m_{1}) A_b (m_{2}) \frac{d
N_{ab}}{d s}. \ee
In this context, the quantity $d N_{q \bar q}/d s$ has the meaning of
the average multiplicity of $q+\bar q$ collisions in the plasma as a
function of the invariant energy of these collisions. Analogously,
$d N_{g q}/ds$ denotes the multiplicity of $g+q$ collisions.

\begin{figure}
\begin{center}
        \resizebox{0.48\textwidth}{!}{%
        \includegraphics{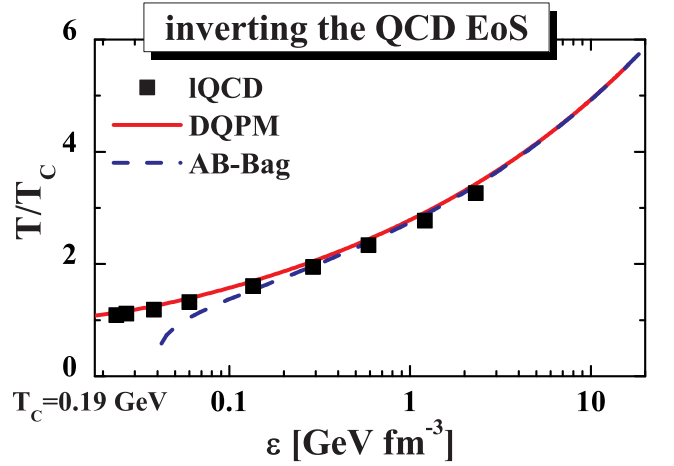}
        }
\caption{(color online) Temperature as a function of the energy
density from lattice QCD
 (square symbols)~\protect{\cite{Karsch:2000ps}}, the
 DQPM approach (red solid line)~\protect{\cite{Cassing:2007nb}} and
 the `AB-Bag' model (blue dashed line)~\protect{\cite{Begun:2010md}}. }
\label{Reverse}
\end{center}
\end{figure}

In order to address the relative importance of the different
mechanisms for the dilepton production in the sQGP, we need a
quantitative model for the multiplicities of the collisions $q+\bar
q$ and $g+q$ ($g+\bar q$) as functions of the center-of-mass energy
$\sqrt{s}$ of these collisions. For this aim, we use the
parametrizations, presented in Fig.~\ref{ModelDNDS}. Inspired by the
Landau model for heavy ion collisions, the parametrizations fall off
with temperature according to the power low $\phi(T)\sim
T^{-7}$~\cite{Shuryak:1978ij}, while the magnitude of the
distributions depend on the total available energy
%($\sim\sqrt{s_{\tiny NN}}\sim <\sqrt{s}>$)
($\sim s_{NN}$). Additionally, we notice that the energy of parton
collisions is on average proportional to the temperature
$\sqrt{s}\sim T$. In this admittedly very simplified picture, the
functional form for $d N/d \sqrt{s}$ is taken as
\be
\frac{dN}{d\sqrt{s}}=N_0 \, s^{1/4}\,\left[ (\sqrt{s}-P)^2+W \right]^{-3.5}.
\ee
%
%qq
%\frac{dN}{d\sqrt{s}}=1900*sqrt(srt)*((srt-0.5)**2+1.2)**(-3.5)
%gq
%\frac{dN}{d\sqrt{s}}=90*sqrt(srt)*((srt-1.2)**2+0.6)**(-3.5)
%
By choosing ($P_{qq}=0.5$, $W_{qq}=1.2$) for $q\bar q$ collisions
and ($P_{gq}=1.2$, $W_{gq}=0.6$) for $gq$ collisions, we adjust
 the maximum of $dN(q+\bar q)$ to $\sqrt{s}\!\approx\!0.5$~GeV
 and shift the maximum of $dN(g+q)$
 to $\sqrt{s}\!\approx\! 1.2$~GeV, reflecting the fact that the threshold
 $\sqrt{s_0}=m_a+m_b$ is higher for $gq$ than for $qq$ collisions.
 Indeed, the gluonic quasi-particles are expected to be more
 massive than the quark ones~\cite{Cassing:2007nb}.
  The parameter $N_0$ remains to be fixed to
experimental data or microscopic calculations, therefore we
currently consider the distributions in ``arbitrary units". At high
$\sqrt{s}$ the distributions approach the Landau ansatz
$\mbox{const} \cdot T^{-7}$, which is shown in Fig.~\ref{ModelDNDS}
by the black dotted line.

In Fig.~\ref{Spectra_with_constant alpha} we plot the dilepton
spectrum, assuming $\alpha_S=0.4$\footnote{The absolute magnitudes
of the dilepton rates presented in Figs.10,14,15 and 16
%~\ref{Spectra_with_constant
%alpha}, \ref{Spectra_with_running alpha}, and \ref{SpectraOff}
 are not to be directly compared to experimental data.
 The analysis of the relative yields serves as an illustration of
an application of the derived off-shell cross sections.}. One
observes that after the convolution with the distribution of
possible $\sqrt{s}$ for the $q+\bar q$ annihilation in sQGP, the
yield of lepton pairs produced in the Bremsstrahlung process is
below the leading order DY mechanism contribution. Thus, the DY rate
is higher in the magnitude than that of the gluon Bremsstrahlung
process despite the fact that the former one contributes only to the
lepton pairs with a mass equal to the $\sqrt{s}$.

On the other hand, one notices from the
Fig.~\ref{Spectra_with_constant alpha} that the GCS mechanism is
sub-leading, unless the gluonic content of the plasma is orders of
magnitude above the quark content, which is achieved neither at SPS
nor at RHIC energies. A very high gluon content
% citation to LHC?
might be found at LHC, in which case, the GCS process would give a considerable
contribution to the dilepton yield of the QGP.

Let us remind that the running coupling $\alpha_S$ depends on the
local energy density $\varepsilon$. The DQPM~\cite{Cassing06}
provides a good parametrization of the QCD running coupling as a
function of temperature in the non-perturbative regime for
temperatures close to $T_c$ (cf. Fig.~1 in Ref.~\cite{Cassing06}).
Note that close
to $T_c$ the full coupling calculated on the lattice increases
with the decreasing temperature much faster than the pQCD prediction.

\begin{figure}
\begin{center}
        \resizebox{0.48\textwidth}{!}{%
        \includegraphics{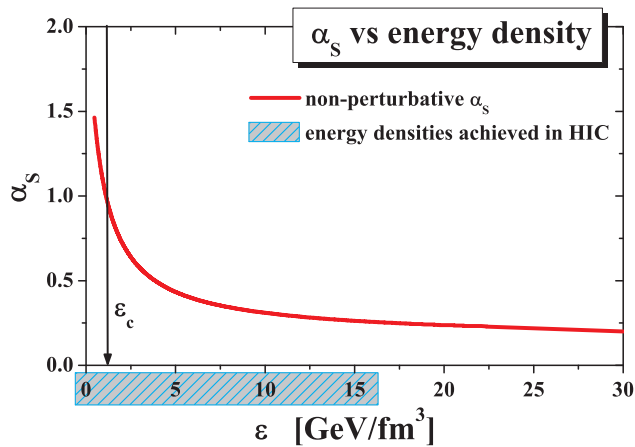}
        }
\caption{(color online) Non-perturbative running coupling as a
function of the local energy density $\varepsilon$. The shadowed
area indicates the energy densities reached in heavy ion collisions
at SPS and RHIC. The arrow shows the critical energy-density.}
\label{alphaSrun}
\end{center}
\end{figure}

The relation between the energy density
and temperature, i.e. the equation of state, has also been extracted
on the lattice~\cite{Karsch:2000ps} -- see Fig.~\ref{EoS}.
The DQMP model describes the QCD equation of state even at $T \sim
T_c$~\cite{Cassing:2007nb}.
A rather simple parametrization for the QCD equation of state -- the
``AB-Bag model'' -- is proposed in~\cite{Begun:2010md} and provides
a good fit of the SU(3) lattice data above $T_c$; we extend this
model to 3-flavors and also compare to lattice
data~\cite{Karsch:2000ps} in Fig~\ref{EoS}.

\begin{figure}
        \resizebox{0.5\textwidth}{!}{%
        \includegraphics{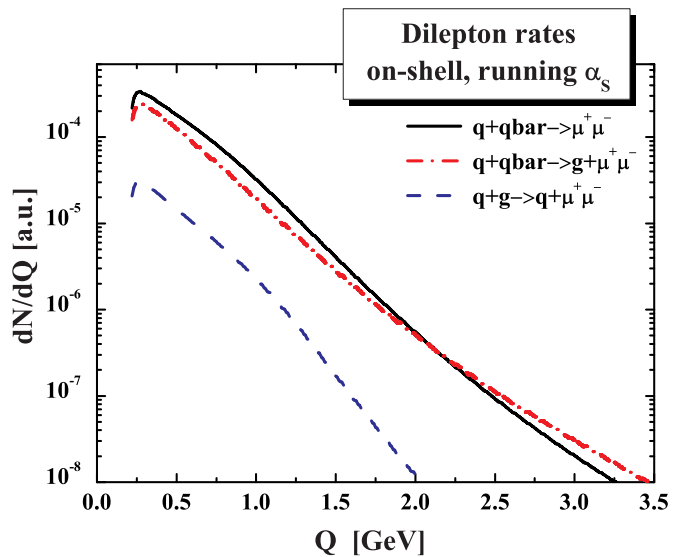}
        }
\caption{(color online) Dimuon rates $d N/dQ$ from QGP calculated
using the cross sections in the on-shell approximation,
$\alpha_S=0.8$. Black solid line shows the contribution of the
Drell-Yan channel ($q+\bar q\to \mu^+\mu^-$), red dash-dotted line
represents the contribution of the channel $q+\bar q\to
g+\mu^-\mu^+$, blue dashed line shows the contribution of the
channel $q+g\to q+\mu^+\mu^-$. The rates are in `arbitrary units',
reflecting our use of simplistic quark and gluon distributions in
the QGP.} \label{SpectraRUN}
\end{figure}

In Fig~\ref{Reverse} we reversed the relation and present the
temperature as a function of the energy density. Using this
relation, we obtain the running coupling as a function of the energy
density $\varepsilon$; we present $\alpha_S$ vs. $\varepsilon$ in
Fig.~\ref{alphaSrun}.
On the other hand, simulations in transport
theory~\cite{Linnyk:2008hp} have shown that the local energy
densities achieved in the course of heavy-ion collisions at SPS and
RHIC energies reach at most $20$~GeV/fm$^3$; this region is
high-lighted in Fig.~\ref{alphaSrun} by a shadowed area.

One observes that $\alpha_S$ at the energy densities of interest is
on average approximately $0.8$. Using this value for $\alpha_S$, we
compare the rates in Fig.~\ref{SpectraRUN}. In this case, the
contribution of the $O(\alpha_S)$ diagrams (gluon-Compton scattering
$qg\to q\gamma^*$ and gluon Bremsstrahlung $q\bar q\to g \gamma^*$)
is no more subleading to the DY annihilation mechanism!

\begin{figure*}
\centering \subfigure[]{
\resizebox{0.48\textwidth}{!}{%
        \includegraphics{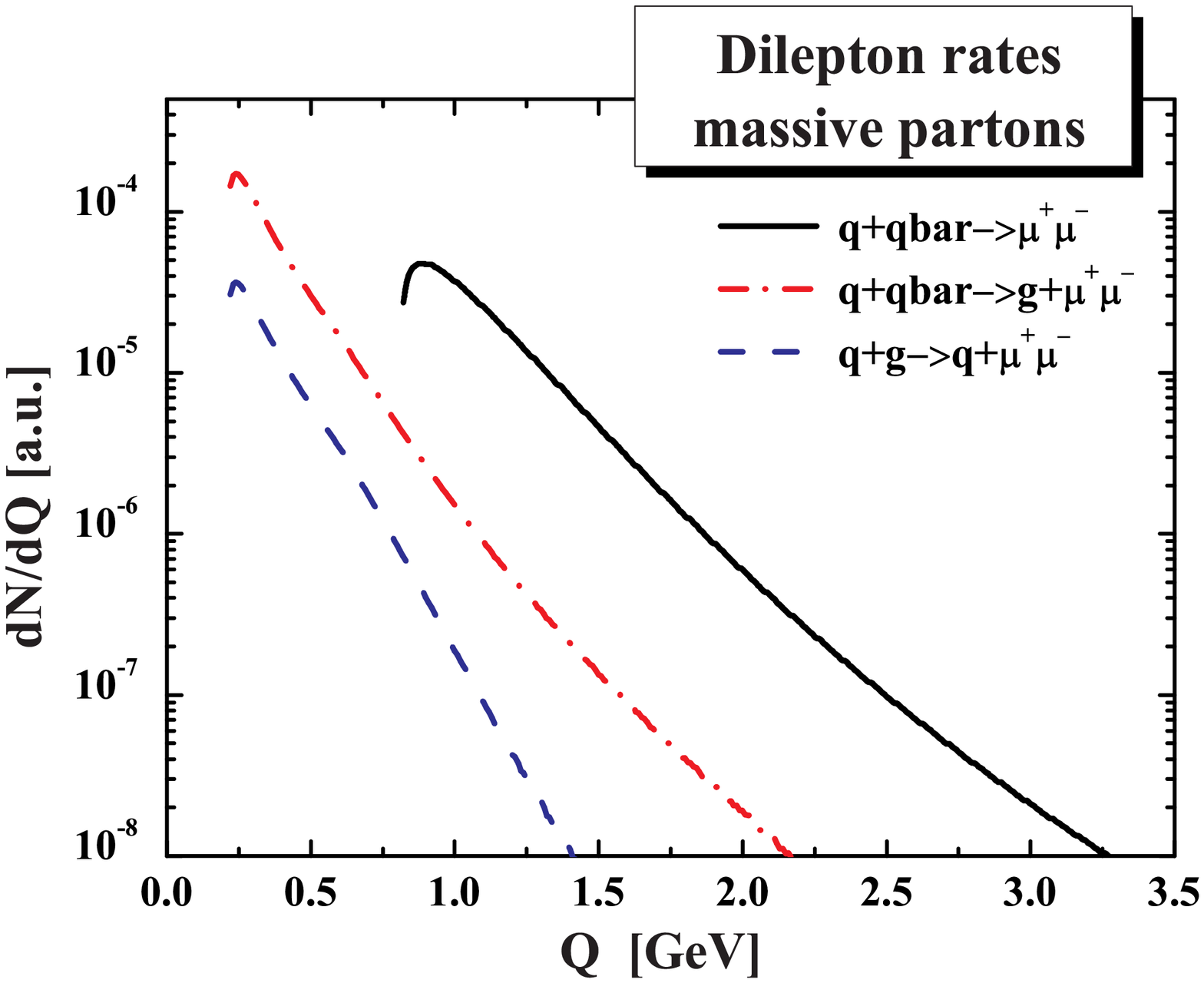}
         } \label{SpectraOff1}
} \subfigure[]{
        \resizebox{0.49\textwidth}{!}{%
        \includegraphics{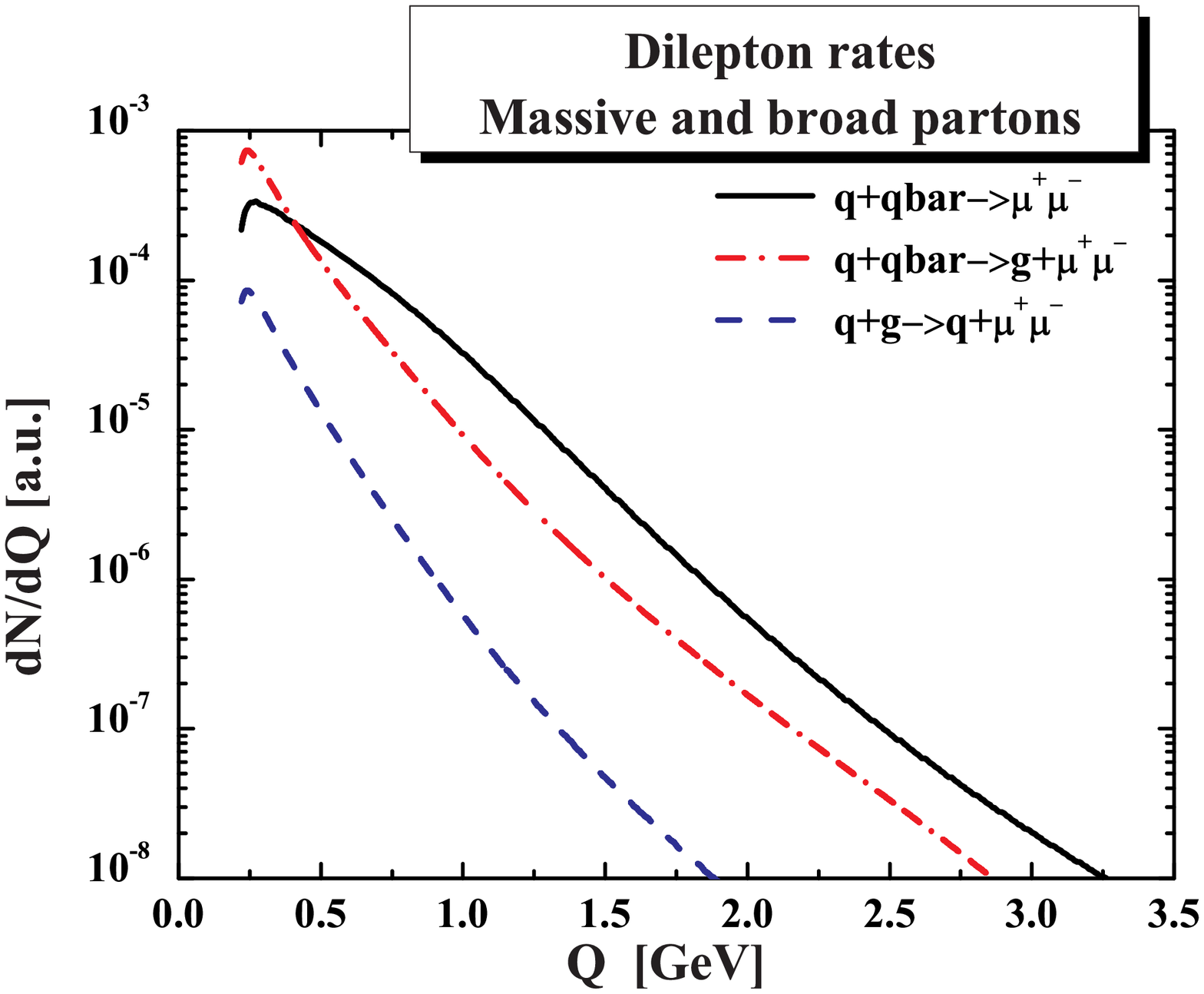}
        } \label{SpectraOff2}
}\caption{(color online) Dimuon rates from QGP $d N/dy$ beyond the
on-shell approximation, $\alpha_S=0.8$.
{\bf L.h.s.} $d N/dQ$ calculated using the derived off-shell cross
sections for quarks and gluons as massive quasi-particles, quark
mass being set to $m_q=0.3$~GeV and gluon mass to $\mu=0.8$~GeV.
Black solid line shows the contribution of the Drell-Yan channel
($q+\bar q\to \mu^+\mu^-$), red dash-dotted line represents the
contribution of the channel $q+\bar q\to g+\mu^-\mu^+$, blue dashed
line shows the contribution of the channel $q+g\to q+\mu^+\mu^-$.
{\bf R.h.s.} $d N/dQ$ calculated in the fully off-shell case of
massive {\em and broad} dynamical quasi-particles. The rates are
calculated using the derived off-shell cross sections and convoluted
with effective spectral functions of the Breit-Wigner type. The
parameters of the spectral functions are: the peak of the quark
spectral function is located at 0.3~GeV, the width is
$\Gamma=0.3$~GeV, the peak of the gluon spectral function is at
0.8~GeV, the width to $\Gamma=0.3$~GeV. Line coding is as on the
l.h.s.
The rates on both plots are in `arbitrary units', reflecting our use
of simplistic quark and gluon distributions in the QGP. }
\label{SpectraOff}
\end{figure*}

Next, we plot the dilepton rates -- within our approximate model for
the parton collision density in the plasma -- for the case of
massive quarks and gluons in Fig.~\ref{SpectraOff1}. The rates are
calculated by convoluting the off-shell cross sections obtained in
the previous Section with model $dN/d\sqrt{s}$. Quarks and gluons
are massive quasi-particles, quark masses being set to $m_q=0.3$~GeV
and gluon mass $\mu=0.8$~GeV. Black solid line shows the
contribution of the Drell-Yan channel ($q+\bar q\to \mu^+\mu^-$),
red dash-dotted line represents the contribution of the channel
$q+\bar q\to g+\mu^-\mu^+$, blue dashed line shows the contribution
of the channel $q+g\to q+\mu^+\mu^-$. The rates in the three
channels are modified compared to the massless case (cf.
Fig.~\ref{SpectraRUN}). In particular, we point to the clear
threshold behavior of the Born (Drell-Yan) term, noted also by the
authors that have studies the dilepton production by massive partons
previously~\cite{Cassing:2007nb,Peshier:1994zf,Thoma:1999nm}.

%%%%%%%%%%%%%%%%%%%%%%%%%%%%%%%%%%%%%%%%%%%%%%%%%%%%%%%%%%%%%%%%%%%%%%%%%%%%%%%%%%
\vspace{10pt}
\section{effect of the final quark and gluon width on the QGP radiation}
\label{effectWidth}

Finally, we calculate the QGP dilepton rate, taking into account not
only the finite masses of the partons, but also their { \em broad
spectral functions}, i.e. finite widths. For this purpose, we
convolute the off-shell cross sections obtained in
section~\ref{offshellsection} with $dN/d\sqrt{s}$ and with the
spectral functions $A(m)$ according to the equation
(\ref{factorization}).

The partonic spectral functions are related to the imaginary part of
the trace of the effective propagator $D_\mu^\mu$ and to the
partonic self-energies $\Sigma$ as follows:
\bea
A(p) \! & \! = \! & \! \frac{1}{\pi} \Im D_\mu^\mu (p) \nn \! & \! =
\! & \! - \frac{1}{\pi} \frac{\Im \Sigma
(p)}{[p^2-m_{current}^2-Re\Sigma (p)]^2+[\Im \Sigma (p)]^2}. \, \,
\eea
For the current qualitative analysis, we use the approximation of
constant real and imaginary parts of the self-energy, which
corresponds to constant finite average mass for quarks ($<m_q>$) and
gluons ($\mu$) and their width $\Gamma$. Within these
approximations, the spectral function has the Breit-Wigner form.

The results of the numeric convolution of the off-shell cross
sections with the spectral functions and the $dN/d\sqrt{s}$ are
shown in Fig.~\ref{SpectraOff2} for realistic values of the spectral
function parameters inspired by DQPM: the peak of the quark spectral
function is located at 0.3~GeV, the width is $\Gamma=0.3$~GeV, the
peak of the gluon spectral function is at 0.8~GeV, the width
$\Gamma=0.3$~GeV.. The rate from the Drell-Yan mechanism is shown by
solid black line, while for the $Q(\alpha_S)$ $2\to 2$ processes by
the dashed blue and dash-dotted red lines. We have checked
numerically that the rates for different values of the gauge
parameter lie on top of each other.
 For comparison, the rates shown on the l.h.s. of Fig.~\ref{SpectraOff}
 have been obtained in the previous Section in the approximation of zero width ($\delta$-functions for
spectral functions).

By dressing the quark and gluon lines with effective spectral
functions we model the effect of the quasi-particle interaction,
including their multiple scattering. One observes by direct
comparison of the l.h.s. and r.h.s. of the Fig.~\ref{SpectraOff}
that the effect on the dilepton rates is quite dramatic. In
particular, the threshold of the Drell-Yan contribution is
``washed-out". In this observation we confirm the results
of~\cite{LPM}. On the other hand, the effect of the partonic width
and/or of multiple scattering on the $2\to2$ processes has not been
studied so far. Weather the predicted shape of the dilepton spectrum
in Fig.~\ref{SpectraOff2} is realized remains to be answered in the
comparison to experimental data~\cite{progress}.

%%%%%%%%%%%%%%%%%%%%%%%%%%%%%%%%%%%%%%%%%%%%%%%%%%%%%%%%%%%%%%%%%%%%%%%%%%%%%%%%%%
\vspace{10pt}
\section{Summary and Outlook}
\label{conclusions}

In the present work, we have derived the off-shell cross sections of
dilepton production by the constituents of the strongly interacting
quark-gluon-plasma in the reactions $q+\bar q\to l^+l^-$ (Drell-Yan
mechanism), $q+ \bar q\to g+l^+l^-$ (quark annihilation with the
gluon Bremsstrahlung in the final state), $q(\bar q)+g\to q(\bar q)+
l^+l^-$ (gluon Compton scattering), $g\to q+\bar q+l^+l^-$ and
$q(\bar q)\to q(\bar q)+g+l^+l^-$ (virtual gluon decay, virtual
quark decay) by dressing the quark and gluon lines in the
perturbative diagrams with effective non-perturbative propagators.

We show that finite quark and gluon virtualities  modify the
magnitude, $Q$ and $q_T$ dependence of the cross sections of all the
calculated processes compared to the leading twist perturbative
results for massless partons. The modification is higher at low
$Q^2$ and on the edges of the phase space.

In QGP, quarks and gluons quasiparticles have broad spectral
functions (due to large interaction rates), where the width encodes
the effects of rescattering and gluon radiation. We have found that
the finite width of partons has a dramatic effect on the dilepton
rates in QGP, especially at low $Q^2$. In particular, the threshold
of the Born term is smeared, while the contribution of the $2\to2$
processes to the rate is increased.

The cross sections obtained in this study will form the basis of a
consistent calculation of the dilepton production in heavy ion
collisions at SPS and RHIC energy by implementing the partonic
processes into a transport approach of the PHSD
collaboration~\cite{progress}. The comparison to the dilepton data
of the NA60 and PHENIX Collaborations double differentially in mass
and $p_T$ will open the possibility to study the relative importance
of different processes in the dilepton production and guide us
towards a better understanding of the properties of matter created
in heavy-ion collisions.

%%%%%%%%%%%%%%%%%%%%%%%%%%%%%%%%%%%%%%%%%%%%%%%%%%%%%%%%%%%%%%%%%%%%%%%%%%%%%%%%%%
\vspace{10pt}
\section*{Acknowledgements}

The author acknowledges valuable discussions with E.~Bratkovskaya,
W.~Cassing, J.~Manninen, V.~Begun and A.~Toia and the financial
support through the ``HIC for FAIR" framework of the ``LOEWE"
program.

%\bibliographystyle{h-physrev3}
%\bibliography{PHSDdilept}

\end{document}